%% file: main.tex
\documentclass[final,5p,times,twocolumn]{elsarticle}
\usepackage{etex}
\usepackage{color}
\usepackage{colortbl}
\usepackage{lscape}
\usepackage{amsmath}
\usepackage{txfonts}
\usepackage{marvosym}
\usepackage[lined,boxed,linesnumbered]{algorithm2e}
\usepackage[ansinew]{inputenc}
\usepackage[french, english]{babel}
\usepackage[T1]{fontenc}

\usepackage{lmodern}  \normalfont
\usepackage{hyperref}
\usepackage{algorithmic}
\algsetup{linenosize=\footnotesize\ttfamily}

\usepackage{amsfonts}
\usepackage{amssymb}
\usepackage{ae,aecompl}
\usepackage{epsfig}

\usepackage{booktabs}
\usepackage{graphics}
\usepackage{array}
\usepackage{graphicx}
\usepackage{tikz}

\usepackage{tabularx}
\usepackage{listings}
\usepackage{longtable}
\usepackage{lineno}

\usepackage{mathrsfs}
\usepackage{lettrine}
\usepackage{tabularx}
\usepackage{epsfig, floatflt, amssymb} 
\usepackage{moreverb} 
\usepackage{cases}  
\usepackage{multirow} 
\usepackage{url} 
\usepackage[all]{xy} 
\usepackage{textcomp} 
\usepackage{setspace} 
\usepackage{epic,eepic}
\usepackage{soul}
\usepackage[nottoc]{tocbibind} 
\usepackage{appendix}
\usepackage{subfig}
\usepackage{ifthen}
\usepackage{xkeyval}
\usepackage{xfor}
\usepackage{amsgen}
\usepackage{etoolbox}
\usepackage{longtable} 
\usepackage{supertabular}
\usepackage{datatool}

\usepackage{fancybox}
\usepackage[leftcaption]{sidecap}
\usepackage[labelsep=endash, textfont={footnotesize, singlespacing}, margin=10pt, format=plain, labelfont=bf]{caption}
\usepackage[Conny]{fncychap} 

\definecolor{name}{rgb}{0.5,0.5,0.5}
\definecolor{javared}{rgb}{0.6,0,0} 
\definecolor{javagreen}{rgb}{0.25,0.5,0.35} 
\definecolor{javapurple}{rgb}{0.5,0,0.35} 
\definecolor{javadocblue}{rgb}{0.25,0.35,0.75} 

\newcommand{\fakeparagraph}[1]{\smallskip\noindent\textbf{#1.}}

\newcommand*\circled[1]{\tikz[baseline=(char.base)]{
            \node[shape=circle,draw,inner sep=0.5pt, minimum size=0.5pt] (char) {#1};}}

\rmfamily
\DeclareFontShape{T1}{lmr}{bx}{sc}{<->ssub * cmr/bx/sc}{}
\DeclareSymbolFont{calletters}{OMS}{cmsy}{b}{n}
\DeclareSymbolFontAlphabet{\mathcal}{calletters}
\DeclareSymbolFont{rmletters}{OMS}{ptm}{m}{n}
\DeclareSymbolFontAlphabet{\mathrm}{rmletters}

\usepackage{mathrsfs}
\usepackage{lettrine}
\usepackage{tabularx}
\usepackage{epsfig, floatflt, amssymb} 
\usepackage{moreverb} 
\usepackage{cases}  
\usepackage{multirow} 
\usepackage{url} 
\usepackage[all]{xy} 
\usepackage{textcomp} 
\usepackage[right]{eurosym}
\usepackage{setspace} 
\usepackage{epic,eepic}
\usepackage{soul}
\usepackage[nottoc]{tocbibind} 
\usepackage{subfig}
\usepackage{ifthen}
\usepackage{xkeyval}
\usepackage{xfor}
\usepackage{amsgen}
\usepackage{etoolbox}
\usepackage{longtable} 
\usepackage{supertabular}
\usepackage{datatool}

\usepackage{fancybox}
\usepackage[leftcaption]{sidecap}

\usepackage[labelsep=endash, textfont={footnotesize, singlespacing}, margin=10pt, format=plain, labelfont=bf]{caption}
\usepackage[Conny]{fncychap} 

\makeatletter
\renewcommand{\thesection}{\arabic{section}}
\renewcommand{\SC@figure@vpos}{c}
\renewcommand{\fnum@figure}{\small\textbf{Figure~\thefigure}}
\renewcommand{\fnum@table}{\small\textbf{Table~\thetable}}

\journal{Journal of Systems and Software}

\begin{document}
\renewcommand\contentsname{Table of Contents}
\renewcommand\listfigurename{List of Figures}
\renewcommand\listtablename{List of Tables}
\renewcommand\bibname{References}
\renewcommand\indexname{Index}
\renewcommand\figurename{Figure}
\renewcommand\tablename{Table}
\renewcommand\partname{Part}
\renewcommand\chaptername{Chapter}
\renewcommand\appendixname{Appendix}

\begin{frontmatter}

\author[rvt]{Pankesh Patel}
\ead{pankesh.patel@ahduni.edu.in}
\author[focal]{Damien Cassou}
\ead{damien.cassou@inria.fr}

\address[rvt]{ABB Corporate Research, India}
\address[focal]{Inria Lille-Nord Europe, France}

\title{Enabling High-level Application Development for the Internet of Things}

\begin{abstract}
\input{abstract}
\end{abstract}

\end{frontmatter}
\input{intro}
\input{ourapproach}
\input{component}

\input{evaluation}

\input{relatedwork}

\input{conclusion}

\section*{Acknowledgment}
This research work is partially done at University of Paris and French National Institute for Research 
in Computer Science  \& Automation (INRIA), France during Pankesh Patel's Ph.D. thesis.
The authors would gratefully like to thank researchers at Inria, Dimitris Soukaras, 
and the reviewers for their helpful comments and suggestions. 

\bibliographystyle{elsarticle-harv}
\bibliography{references}

\end{document}

%% file: abstract.tex
Application development in the Internet of Things (IoT) is challenging because it involves dealing with
a wide range of related issues such as lack of separation of concerns, and lack of high-level of abstractions
to address both the large scale and heterogeneity. Moreover, stakeholders involved in the application development
have to address issues that can be attributed to different life-cycles phases. when developing applications.
First, the application logic has to be analyzed and then separated into a set of distributed tasks for an underlying network.
Then, the tasks have to be implemented for the specific hardware. Apart from handling these issues, they have
to deal with other aspects of life-cycle such as changes in application requirements and deployed devices.

Several approaches have been proposed in the closely related fields of wireless sensor network, ubiquitous and pervasive
computing, and software engineering in general to address the above challenges. However, existing approaches only cover limited
subsets of the above mentioned challenges when applied to the IoT. This paper proposes an integrated approach for addressing the above
mentioned challenges. The main contributions of this paper are: (1) a development methodology that separates
IoT application development into different concerns and provides a conceptual framework to develop an application,
(2) a development framework that implements the development methodology to support actions of stakeholders. The
development framework provides a set of modeling languages to specify each development concern and abstracts
the scale and heterogeneity related complexity. It integrates code generation, task-mapping, and linking techniques
to provide automation.  Code generation supports the application development phase by producing a programming
framework that allows stakeholders to focus on the application logic, while our mapping and linking techniques
together support the deployment phase by producing device-specific code to result in a distributed system
collaboratively hosted by individual devices. Our evaluation based on two realistic scenarios shows that
the use of our approach improves the productivity of stakeholders involved in the application development.


%% file: intro.tex
\section{Introduction}
The recent technological advances have been fueling a tremendous growth in a number of smart objects 
~\cite[p.~3]{vasseur2010interconnecting} such as temperature sensors, 
smoke detectors, fire alarms, parking space controllers. They can sense the physical world 
by obtaining information from sensors, affect the physical world by triggering actions 
using actuators, engage users by interacting with them whenever necessary, and process 
captured data and communicate it to outside world. In the \emph{Internet of Things}~\cite{casagras}, 
smart objects~(or ``things'') acquire intelligence 
thanks to the fact that they can communicate with each other and cooperate with their neighbors 
to reach a common goal~\cite{IoTsurvey}.  For example, a building interacts with 
its residents and surrounding buildings in case of fire for safety and security of residents, offices 
adjust themselves automatically accordingly to user preferences while minimizing energy consumption, 
or traffic signals control in-flow of vehicles according to the current highway status~\cite{de2009internet}.

As evident above, IoT applications will involve interactions among large numbers of disparate devices, 
many of them directly interacting with their physical surroundings. An important challenge that needs 
to be addressed in the IoT, therefore, is to enable the rapid development of IoT applications with 
minimal effort by the various stakeholders\footnote{Throughout this paper, we use the term \textbf{stakeholders} 
as used in software engineering to mean -- people, who are involved in the application development. Examples of 
stakeholders defined in \cite{softwareArchtaylor2010} are software designer, developer, domain expert, 
technologist, etc.} involved in the process. Similar challenges have already been addressed in the closely 
related fields of Wireless Sensor Networks (WSNs)~\cite[p.~11]{vasseur2010interconnecting} and ubiquitous 
and pervasive computing~\cite[p.~7]{vasseur2010interconnecting}, regarded as precursors to the modern day IoT.  
While the main challenge in the former is the \emph{large scale} -- hundreds to thousands of largely similar 
devices,  the primary concern in the latter  has been the {\em heterogeneity} of devices and the major role 
that the  user's own interaction with these devices plays in these systems (cf. the classic ``smart home'' 
scenario where a user controls lights and receives notifications from  his  refrigerator and toaster.).  
It is the goal of our work to enable the development of such applications. In the following, 
we discuss one of such applications.

\subsection{Application example}\label{sec:appexample}
\begin{figure}[!ht]
\centering
\includegraphics[width=0.45\textwidth]{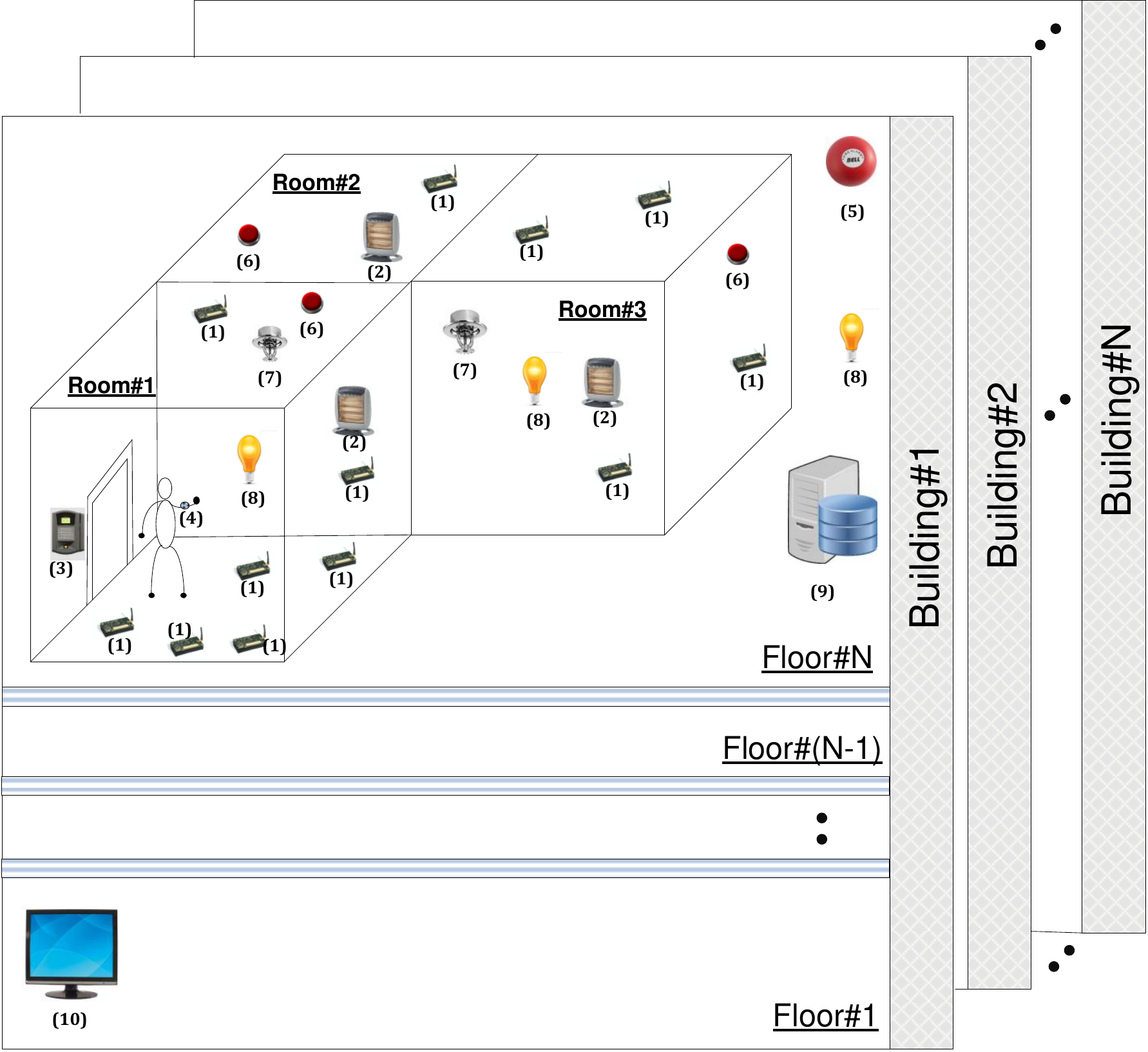}
\caption{A cluster of multi-floored buildings with deployed devices with (1)~temperature sensor, 
 (2)~heater, (3)~badge reader, (4)~badge, (5)~alarm, (6)~smoke detector, (7)~sprinkler, (8)~light, (9)~data storage, and 
 (10)~monitor.}
\label{fig:casestudybuilding}
\end{figure}

We consider a hypothetical building system utilized by a company. This building system might 
consist of several buildings, with each building in turn consisting of one or more floors, 
each with several rooms. It may consist of a large number of heterogeneous devices equipped with sensors, actuators, 
storage, user interfaces. Figure~\ref{fig:casestudybuilding} describes the building automation 
domain with various devices. Many applications can be developed using these devices, one of 
which we discuss below.

\fakeparagraph{Smart building application}
To accommodate the mobile worker's preference in the reserved room, a database is used to keep the profile 
of each worker, including his preferred lighting and temperature level.  
A badge reader in the room detects the worker's entry event and queries the database for the worker's 
preference. Based on this, the thresholds used by the room's devices are updated.  To reduce electricity 
waste when a person leaves the room, detected by badge disappeared event, lighting and heating 
level are automatically set to the lowest level; all according to the building's policy. 
The system may also include user interfaces that allow a late worker to control heater of his room 
and request the profile database to get his lighting and temperature preferences. Moreover, the system 
generates the current status (e.g., temperature, energy consumption) of each room, 
which is then aggregated and used to determine the current status of each floor and, in turn, 
the entire building. A monitor installed at the building entrance presents the information 
to the building operator for situational awareness. 

\subsection{IoT application development challenges}\label{sec:challenges}

This section reviews the application development challenges as gleaned from our analysis of 
applications such as the one discussed in the previous section.  The challenges we address in 
this work are as follows:

\fakeparagraph{\emph{Lack of division of roles}}
IoT application development is a multi-disciplined process where 
knowledge from multiple concerns intersects. Traditional IoT application 
development assumes that the individuals involved in the application development 
have similar skills. This is in clear conflict with the varied set of skills 
required during the process, including domain 
expertise~(e.g., the smart 
		building application reason in terms of rooms and floors, the smart city applications are expressed 
		in terms of sectors.), deployment-specific knowledge~(e.g., understanding of the specific target area where the application is to be deployed, 
     mapping of processing components to devices in the target deployment),  application design and implementation knowledge, 
 and platform-specific knowledge~(e.g., Android-specific APIs  to get data 
		from sensors, vendor-specific database such as MySQL), a challenge recognized by recent works 
such as~\cite{towardchen2012, softwarepicco2010}.

\fakeparagraph{\emph{Heterogeneity}} 
IoT applications execute on a network consisting of heterogeneous 
devices  in terms of types~(e.g., sensing, actuating, storage, and user interface devices), interaction 
modes~(e.g. Publish/Subscribe~\cite{eugster2003many}, Request/Response~\cite{berson1996client}, 
Command~\cite{andrews1991paradigms}), as well as different 
platforms~(e.g., Android mobile OS, Java SE on laptops). The heterogeneity largely spreads into the application 
code and makes the portability of code to a different deployment 
difficult.  

\fakeparagraph{\emph{Scale}} 
As mentioned above, IoT applications execute on distributed systems consisting of hundreds to 
thousands of devices, involving the coordination of their activities~(e.g., temperature values are computed at per-room and then 
per-floor levels to calculate an average temperature value of a building). Requiring the ability of reasoning at such levels 
of scale is impractical in general, as has been largely the view in the WSN community.

\fakeparagraph{\emph{Different life cycle phases}}
Stakeholders have to address issues that are attributed to different life cycles phases,including development, deployment, and maintenance~\cite{bischoff2007life}. 
At the \textbf{development phase}, the application logic has to be analyzed and separated into a set of distributed 
tasks for the underlying network consisting of a large number of heterogeneous entities. 
Then, the tasks have to be  implemented for  the specific platform of a device. At the \textbf{deployment 
phase}, the application logic has to be deployed onto a large number of devices.  Apart from handling these 
issues, stakeholders have to keep in mind evolution issues both in the development (change in functionality 
of an application such as  the smart  building application is  extended by including fire detection functionality) 
 and deployment phase (e.g. adding/removing devices in deployment scenarios such as  more temperature sensors are added  to sense accurate  
temperature values in the building) at the \textbf{maintenance 
phase}.  Manual effort in all above three phases for hundreds to thousands of heterogeneous devices is 
a time-consuming and error-prone process. 

In order to address the above mentioned challenges, various approaches have been proposed (for a
detailed discussion of various systems available for application development, refer Section~\ref{chpt:relatedwork}).
One of the approach is node-centric programming~\cite{hoodwhitehouse2004, roman2002gaia, tinylimecosta2007}. It allows for the development 
of extremely efficient systems based on complete control over individual devices. 
However, it is not easy to use for IoT applications due to the large size and heterogeneity of systems.
In order to address node-centric programming limitation, various 
macroprogramming systems~\cite{pathak2007compilation,bischoff2007life} have been proposed.
However, most of macroprogramming systems largely focus on development phase while ignoring the fact that it 
represents a tiny fraction of the application development life-cycle. 
The lack of a software engineering methodology to support the entire application development life-cycle commonly 
results in highly difficult to maintain, reuse, and platform-dependent design, which can be tackled by 
the model-driven approach. To address the limitations of macroprogramming systems, approaches based on
model-driven design~(MDD) have been proposed~\cite{intromde, france2007model, 
mellor2003model, kulkarni2003separation}. Major benefits came from the basic idea that by separating 
different concerns of a system  at a certain level of abstraction,  and  by providing transformation 
engines to convert  these abstractions to a target code, productivity~({\em e.g.}, reusability, maintainability) 
in the application development process can be improved.

\subsection{Contributions}\label{sec:contri}

Our aim is to make IoT application development easy  for stakeholders as is the case in software 
engineering in general, by taking inspiration from the MDD approach and building upon work in 
sensor network macroprogramming. We achieve this aim by  
separating IoT application development into different concerns and integrating a set of high-level 
languages\footnote{Please note that high-level languages (e.g., AADL, EAST-ADL, SysML, etc.) 
for IoT have been investigated at length in the domains of  pervasive/ubiquitous  
computing and wireless sensor network. However, their integration to our development framework 
in an appropriate way is our contribution.} to specify them.  We provide 
automation techniques at different phases of IoT application development to reduce development effort. 
We now present these contributions in detail described below:

\fakeparagraph{\emph{Development methodology}} We propose a development methodology 
that defines a precise sequence of steps to be followed to develop IoT applications, thus 
facilitating IoT application development. These steps are separated into four 
concerns, namely, domain, functional, deployment, and platform. This separation allows 
stakeholders to deal with them individually and reuse them across applications. 
Each concern is matched with a precise stakeholder according to skills. The clear 
identification of expectations and specialized skills of each type of stakeholders 
helps them to play their part effectively. 

\fakeparagraph{\emph{Development framework}} To support the actions of each stakeholder, the development 
methodology is implemented as a concrete development framework\footnote{It includes support programs, 
   code libraries, high-level languages or other software that help stakeholders to develop and glue together 
	 different components of a software product.}. It provides a set of modeling languages, 
each named after ``Srijan'',\footnote{{\em Srijan} is the sanskrit word for ``creation''.} 
and offers automation techniques at different phases of IoT application development, including the following: 

\begin{itemize} 
\item \fakeparagraph{\emph{A set of modeling languages}} To aid stakeholders, the development 
framework integrates three modeling languages that abstract  the scale and heterogeneity-related 
complexity: (1) Srijan Vocabulary Language~(SVL) to describe domain-specific features  of an IoT 
application, (2) Srijan Architecture Language~(SAL) to describe application-specific  functionality 
of an IoT application, (3) Srijan Deployment Language~(SDL) to describe deployment-specific features  
consisting information about a physical environment where devices are deployed. 

\item \fakeparagraph{\emph{Automation techniques}} The development framework is supported by code-generation, 
task-mapping, and linking techniques.  These three techniques together provide automation at 
various phases of IoT application development.  Code generation supports the application development 
phase by producing a programming framework that reduces the effort in specifying the details of the components 
of an IoT application. Mapping and linking together support the deployment phase by producing 
device-specific code to result in a distributed system collaboratively hosted by individual devices.
\end{itemize}

Our work on the above is supported at the lower layers by a middleware that enables delivery of messages 
across physical regions, thus enabling our abstractions for managing large scales in the Internet of Things.

\fakeparagraph{Outline} 
The remainder of this paper is organized as follows:
Section~\ref{chapt:ourapproach} presents our development methodology and its  development framework. 
This includes details of on modeling languages, automation techniques, and our approach for handling evolutions.
Section~\ref{sec:components} presents an implementation of our development framework. We present tools, 
technologies, and programming languages used to implement this development framework. 
Section~\ref{chpt:evolution} evaluates the development framework in a quantitative manner.
Section~\ref{chpt:relatedwork} explores state of the art approaches for developing IoT applications.
Section~\ref{chapt:conclusionandfuturework} summarizes this paper  
and  Section~\ref{sec:futurework} describes briefly some future directions of this work.

%% file: ourapproach.tex
\section{Our approach to IoT application development}\label{chapt:ourapproach}

Applying separation of concerns design principal from software engineering, we 
break the identified concepts and associations among them into different concerns 
represented in Conceptual model~\cite{appdevIoTpatel2011}, described in Section~\ref{sec:conceptualmodel}. The identified concepts are 
linked together into a well-defined and structured methodology, described in Section~\ref{sec:methodology}. 
We  implement the proposed development methodology as a concrete development 
framework~\cite{patelhal00809438, patel-icse14, patel-comnet-iot15,patel-thesis14},
presented in Section~\ref{sec:detail-ourapproach}.

\subsection{Conceptual model}\label{sec:conceptualmodel}
\begin{figure*}[!ht]
	\centering \includegraphics[width=0.65\textwidth]{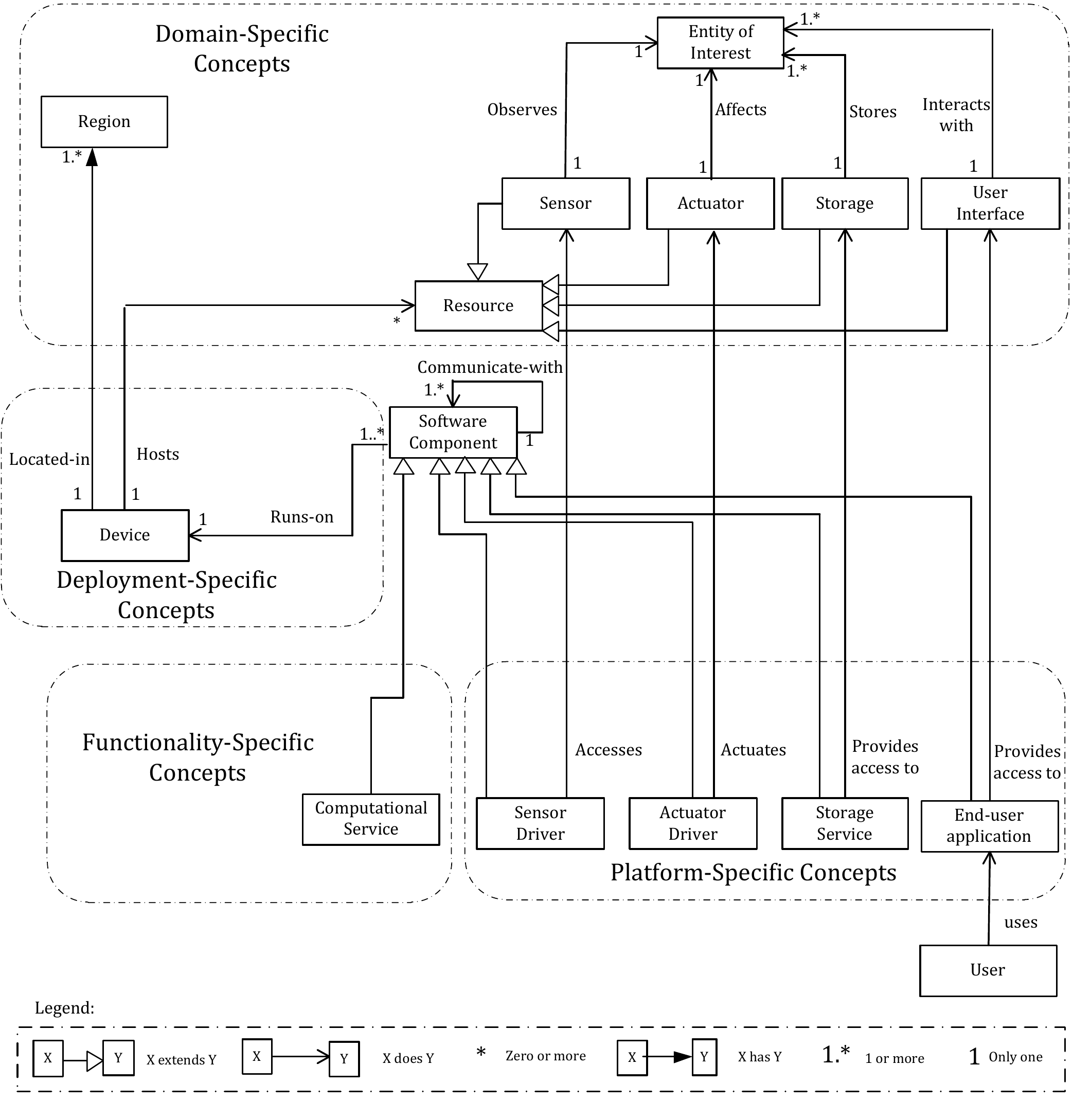} 
	\caption{Conceptual model for IoT applications} \label{fig:conceptualmodel}
\end{figure*}

A conceptual model often serves as a base of knowledge about a problem 
area~\cite{fowler1996analysis}.  It represents the concepts as well as the associations  
among them and also attempts to clarify the meaning of various terms. 
Taking inspiration from previous efforts~\cite{bischoff2007life,towardscassou2011,doddapaneni2012model}, 
we have identified four major concerns for IoT application development. Figure~\ref{fig:conceptualmodel} 
illustrates the concepts and their associations along with these four separate concerns:  
(1) domain-specific concepts, (2) functionality-specific concepts, (3) deployment-specific concepts, 
and (4) platform-specific concepts.

\subsubsection{Domain-specific concepts}
The concepts that fall into this category are specific to a target application 
domain~(e.g., building automation, transport, etc.). For example, the building automation domain is
reasoned  in terms of rooms and floors, while the transport domain is expressed in terms of highway sectors. 
Furthermore, each domain has a set of entities of interest~(e.g., average temperature of a building, 
smoke presence in a room), which are observed and controlled by sensors and actuators respectively. Storages store information 
about entities of interest, and user interfaces enable users to interact with entities of interest (e.g., receiving notification in case of fire in a building, controlling the temperature of a room). We describe these concepts in detail below:

\begin{itemize}
 
\item An \textbf{Entity of Interest~(EoI)} is an object~(e.g., room, book, plant), including  
attributes that describe it, and its state that is relevant from 
a user or an application perspective~\cite[p.~1]{thingshaller2010}. The entity of interest has an observable property 
called {\em phenomenon}. Typical examples are the temperature value of a room and a tag ID.    

\item A \textbf{resource} is a conceptual representation of a
sensor, an actuator, a storage, or a user interface. We consider the following 
types of resources:

\begin{itemize}
\item A \textbf{sensor} has the ability to detect changes in the
			environment. Thermometer and tag readers are examples of
			sensors. The sensor \textbf{observes} a phenomenon of an EoI. For instance,
			a temperature sensor observes the temperature phenomenon of a room.			
\item An \textbf{actuator} makes changes in the environment through an action. Heating or cooling
		elements, speakers, lights are examples of actuators. 
		The actuator \textbf{affects} a phenomenon of an EoI by performing
		actions. For instance, a heater is set to control a temperature level of a room.					
		
	\item A {\bf storage}  has the ability of  storing  data in a persistent manner. The storage \textbf{stores} information about a phenomenon of an EoI. 
	For instance, a database server stores information about an employee's temperature  
	preference.

	 \item A \textbf{user interface} represents tasks available to users to interact with  
	       entities of  interest. For  the building automation domain, a task could be receiving a 
				fire notification in case of emergency or controlling a heater according to a temperature preference.

\end{itemize}

\item 
A  device is located in  a {\bf region}~\cite{tubaishat2003sensor}. The region is used to specify the location
            of a device. In the building automation domain, a region (or location) of a 
						device can be expressed in terms of building, room, and floor IDs.

\end{itemize}

\subsubsection{Functionality-specific concepts}
The concepts that fall into this category describe computational elements of an application
and interactions among them. A computational element is a type of  software component, which is an
architectural entity that (1)~encapsulates a subset of the 
system's functionality and/or data, (2)~restricts access to that subset 
via an explicitly defined interface~\cite[p.~69]{softwareArchtaylor2010}. We use the term \textbf{application logic}
to refer a functionality of a software component. An example of the application logic 
is to open a window when the average temperature value of a room is greater than $30^{\circ} C$. 

The conceptual model contains the following functionality-specific software component, a {\bf computational service}, which is a type of software component that 
			consumes one or more units of information as inputs, processes it, and 
			generates an output. An output could be data message that is consumed by others or a command message that triggers an action of an actuator.
			A computational service is a representation of the processing element in an application. 				

A software component \textbf{communicates-with} other 
software components to exchange data or control. These interactions might contain instances of 
various interaction modes such as request-response, 
publish-subscribe, and command. Note that this is in principle an instance of the component-port-connector architecture used in software engineering.

\subsubsection{Deployment-specific concepts}
The concepts that fall into this category describe information about devices. 
Each device \textbf{hosts} zero or more resources. For example, a device could host resources such as 
a temperature sensor to sense, a heater to control a temperature level, a monitor to display a temperature value, 
a storage to store temperature readings, etc. Each device is \textbf{located-in} regions. 
For instance, a device  is located-in room\#1 of floor\#12 in building\#14.
We consider the following definition of a device:

\begin{itemize}
	
		\item A \textbf{device} is an entity that provides resources the ability of interacting with other devices. 
		Mobile phones, and personal computers are examples of devices.
\end{itemize}

\subsubsection{Platform-specific concepts}
The concepts that fall into this category are computer programs that act as a (operating system-specific) translator between 
a  hardware device and an application. 
We identify the following platform-specific concepts:

\begin{itemize}
	\item A {\bf sensor driver} is a type of software component that operates on a sensor attached to a device. 
	It \textbf{accesses} data observed by the sensor and generates the meaningful
	data that can be used by other software components. 
	For instance, a temperature sensor driver generates temperature values 
  and its meta-data such as unit of measurement, time of sensing. 
	Another software component takes this temperature data as input 
	and calculates the average temperature of the room.

	\item An {\bf actuator driver} is a type of software component that controls an actuator 
	  attached to a device.  It translates a command from other software components and \textbf{actuates} the  
		actuator appropriately. For instance, a heater driver translates a command 
		``turn the heater on'' to regulate  the  temperature level.

	\item A \textbf{storage service} is a type of software component that provides a read and write 
	   access to a storage. 	A storage service \textbf{provides access to} the storage. 
		Other software components access data from the storage by requesting the storage service.  
	 For instance, MySQL storage service provides access to a database server.	
        
    \item An \textbf{end-user application} is a type of software component that is designed to 
		help a user to perform tasks~(e.g., receiving notifications, submitting information). 
		It \textbf{provides access to} available tasks. For instance, in the smart building application 
		a user could provide his temperature 	preferences using an application installed on his smart phone.  

\end{itemize}

The next section presents a development methodology that links 
the above four concerns and provides a conceptual framework to develop IoT applications.

\subsection{A development methodology}\label{sec:methodology}
To make IoT application development easy,  stakeholders should be provided 
a structured and well-defined application development process (referred to as 
{\em development methodology}). This section presents a development methodology 
that integrates  different development concerns discussed in 
Section~\ref{sec:conceptualmodel} and provides a conceptual  framework  
for  IoT application development. In addition to this, it
assigns a precise role to each stakeholder commensurate with his
skills and responsibilities.

As stated in Section~\ref{sec:challenges}, IoT application development is a multi-disciplined 
process where knowledge from multiple concerns intersects. So far, IoT application development 
assumes that the individuals  have similar skills. While this 
may be  true for simple/small applications for single-use deployments, as the IoT gains wide 
acceptance, the need for sound software engineering approaches to adequately manage the 
development of complex applications arises.

\begin{figure*}[!ht]
\centering
\includegraphics[width=0.75\textwidth]{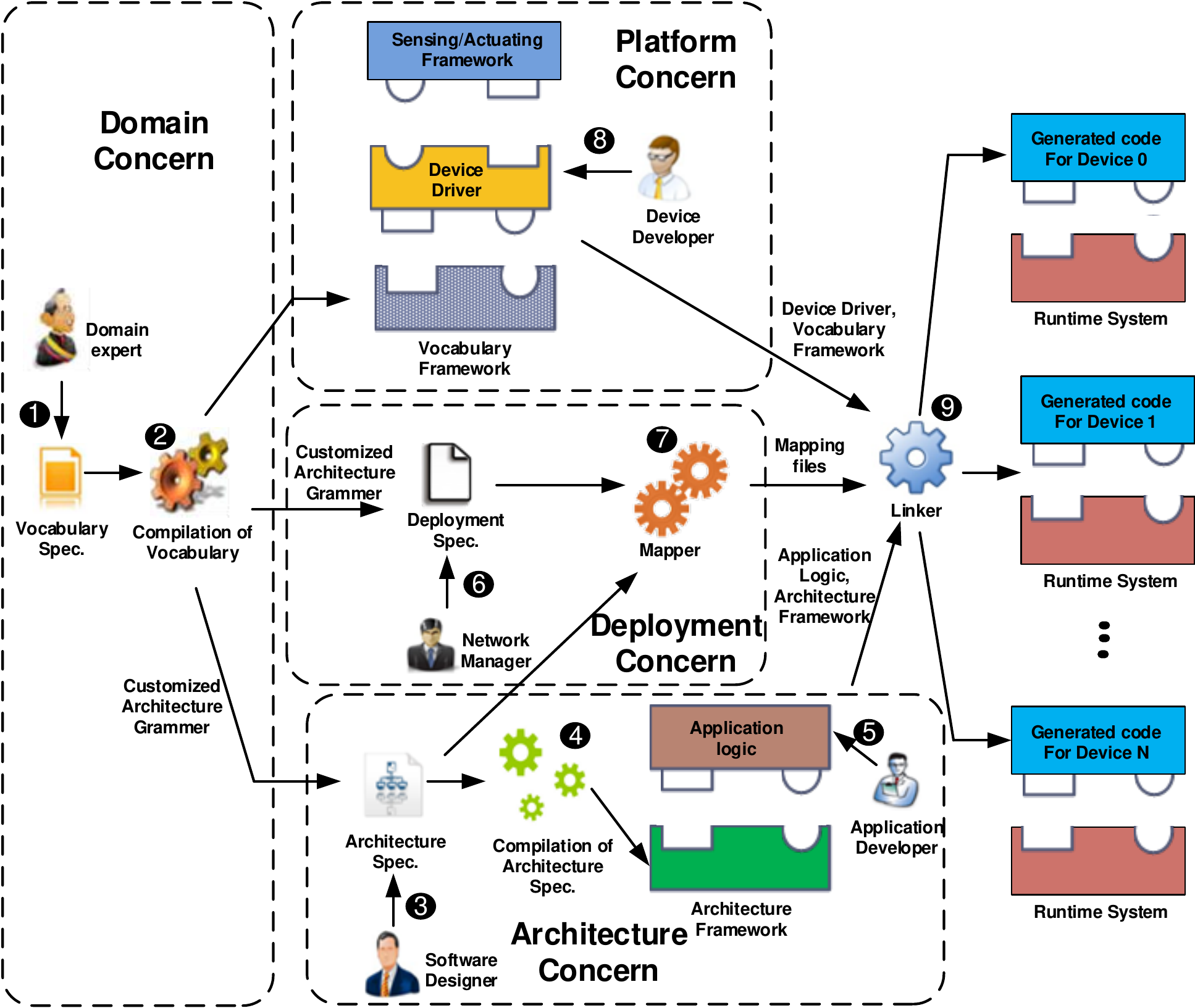}
\caption{IoT application development: overall process}
\label{fig:devcycle}
\end{figure*}

Taking inspiration from ideas proposed in the 4+1 view model of software architecture~\cite{fourplusoneviewkruchten1995}, 
collaboration model for smart spaces~\cite{towardchen2012}, and tool-based methodology 
for pervasive computing~\cite{towardscassou2011}, we propose a development methodology 
that provides a conceptual framework to develop an IoT application~(detailed in Figure~\ref{fig:devcycle}). 
The development methodology divides the responsibilities of stakeholders 
into five distinct roles ---domain expert, software designer, application developer, device developer, 
and network manager.  
Note that although these roles have been discussed in the software engineering literature in general, e.g.,
domain expert and software designer in~\cite[p.~657]{softwareArchtaylor2010}, application 
developer~\cite[p.~3]{towardscassou2011}, their  clear identification for IoT applications is largely missing. Due to the existence of various, slightly varying, definitions in literature, we summarize the skills and responsibilities of the various stakeholders in Table~\ref{tab:roles}

\setlength{\tabcolsep}{3pt}
\begin{table}[htb]
\centering
\footnotesize
\begin{tabular}{ p{0.20\linewidth}  p{0.35\linewidth}  p{0.35\linewidth}  }
 \toprule
{\bf Role} & {\bf Skills} & {\bf Responsibilities} \\ \midrule 
Domain expert & Understands domain concepts, including the data types produced by the sensors, consumed by actuators, accessed from storages, user's interactions, and how the system is divided into regions. 
& Specify the vocabulary of an application domain to be used by applications in the domain.\\ \midrule

Software designer & Software architecture concepts, including the proper use of interaction modes such as
publish-subscribe, command, and request-response for use in the application. &
Define the structure of an IoT application by specifying the software
components and their generate, consume, and command relationships.\\ \midrule

Application developer 
&Skilled in algorithm design and use of
programming languages. & Develop the application logic of the
computational services in the application.\\ \midrule

Device developer & Deep understanding of the inputs/outputs, and protocols of
the individual devices. & Write drivers for the sensors, actuators, storages, and end-user
applications used in the domain.\\ \midrule

Network manager & Deep understanding of the specific
target area where the application is to be deployed. &
Install the application on the system at hand; this process may
involve the generation of binaries or bytecode, and configuring
middleware. \\ \bottomrule

\end{tabular}
\caption{Roles in IoT application development\label{tab:roles}}
\end{table}

An application corresponds to a specific application domain~(e.g., building automation, health-care, transport) 
consisting of domain-specific concepts.  Keeping this in mind, we separate the domain concern from other 
concerns~(see Figure~\ref{fig:devcycle}, \textbf{stage 1}). 
The main advantage of this separation is that domain-specific knowledge can be made 
available to stakeholders and reused across applications of 
a same application domain.

IoT applications closely interact with the physical world. Consequently, changes in either of them have 
a direct influence on the other. The changes could be technological advances with new software features, 
a change in functionality of an application, a change in distribution of devices, and adding or replacing devices. Considering this aspect, we separate IoT application development
into the  platform, functional, and deployment concern at the second stage~(see Figure~\ref{fig:devcycle}, \textbf{stage 2}). 
Thus, stakeholders can deal with them individually and reuse them across applications.  The final stage combines and packs 
the code generated by  the second stage into packages that be deployed on devices~(see Figure~\ref{fig:devcycle}, \textbf{stage 3}).  

\subsection{Development framework}\label{sec:detail-ourapproach}
To support  actions of  stakeholders,  the  development methodology discussed in 
Section~\ref{sec:methodology} is implemented as a concrete development framework. 
This section presents  this development framework that provides a set 
of modeling languages, each named after {\em Srijan}, and offers automation techniques 
at different phases of IoT application development for the respective concerns.

\subsubsection{Domain concern} 
This concern is related to domain-specific concepts of  an  IoT application. 
It consists of the following steps:
\begin{itemize}
\item \fakeparagraph{Specifying domain vocabulary} 
The domain expert specifies a domain vocabulary~(step~\circled{1} in Figure~\ref{fig:devcycle}) using the Srijan Vocabulary
 Language~(SVL). 
The vocabulary includes specification of resources, which are responsible 
for interacting with entities of interest. In the vocabulary, resources  are specified 
in a high-level manner to abstract low-level details from the domain expert.  
Moreover, the vocabulary includes definitions of regions that 
define spatial partitions~(e.g., room, floor, building) of a system. 

\item \fakeparagraph{Compiling vocabulary specification}
    Leveraging the vocabulary, the development framework generates~(step~\circled{2} in Figure~\ref{fig:devcycle}): (1) a vocabulary framework to aid 
    the device developer, (2) a customized architecture grammar according to the vocabulary 
		to aid the software designer, 
		and (3) a customized deployment grammar according to the vocabulary to aid the network manager. 
		The key advantage of this customization  is that the
		domain-specific concepts defined in the vocabulary are made available to other stakeholders and can 
		be reused across applications of the same application domain.

\end{itemize}

\subsubsection{Functional concern}
This concern is related to functionality-specific concepts of an IoT application. It consists of the following steps:
\begin{itemize}
\item \fakeparagraph{Specifying application architecture} 
Using a customized architecture grammar, the software designer 
specifies  an application architecture~(step~\circled{3} in Figure~\ref{fig:devcycle}) using the Srijan Architecture Language~(SAL). 
SAL is an  architecture  description language (ADL)  designed for specifying computational services 
and  their  interactions with other software components. To facilitate scalable operations within IoT applications, 
SAL offers scope constructs. These constructs allow the software designer to group devices based on their 
spatial relationship to form a cluster (e.g., ``devices are in room\#1'') and to place a cluster head to receive 
and process data from that cluster. The grouping and cluster head mechanism 
can be recursively applied to form a hierarchical clustering that facilitates the scalable operations
within IoT applications.

\item \fakeparagraph{Compiling architecture specification}  
The development framework leverages an architecture specification to support 
the application developer~(step~\circled{4} in Figure~\ref{fig:devcycle}). To describe the application logic of
each computational service, the application developer is provided an
architecture  framework, pre-configured according to the
architecture specification of an application, 
an approach similar to the one discussed in~\cite{generativecassou2009}. 
\item \fakeparagraph{Implementing application logic} 
To describe the application logic of each computational service, the application 
developer leverages a generated architecture framework~(step~\circled{5} in Figure~\ref{fig:devcycle}). It contains abstract 
classes\footnote{We assume that  the application developer uses an object-oriented 
language.}, corresponding to  each  computational service, 
that hide interaction details with other software components  and allow the application developer to 
focus only  on application logic. The application developer implements only the abstract 
methods of generated abstract classes.
\end{itemize}

\subsubsection{Deployment concern}
This concern is related to  deployment-specific concepts of an IoT application. 
It consists of the following steps:
\begin{itemize}
\item \fakeparagraph{Specifying target deployment} 
Using a customized deployment grammar, the network manager describes a deployment 
specification~(step~\circled{6} in Figure~\ref{fig:devcycle})
using the Srijan Deployment Language~(SDL).  The deployment specification  includes the details of each 
device\footnote{Our work excludes low-end computing devices (e.g., SunSpoT, TelosB, Tmote Sky, etc.).    We believe this is   a reasonable assumption because technological advances in embedded system result into  devices with more and more computational power and memory.}, including its regions (in terms of values of the regions 
defined in the vocabulary), resources hosted by devices (a subset of those defined 
in the vocabulary), and the type of the device. Ideally, the same IoT application could  
be deployed on different target deployments~(e.g., the same inventory tracking 
application can be deployed in different warehouses). This requirement is 
dictated by separating  a deployment specification  from other specifications.

\item \fakeparagraph{Mapping}
The mapper produces a mapping from a set of
computational services to a set of devices~(step~\circled{7} in Figure~\ref{fig:devcycle}). 
It takes as input a set of placement rules of computational services from an architecture 
specification and a set of devices defined in a deployment specification.  
The mapper decides devices where each computational service will be deployed.

\end{itemize}

\subsubsection{Platform concern}

This concern is related to platform-specific concepts of an IoT application. 
It consists of the following step:
\begin{itemize}  
    \item \fakeparagraph{Implementing device drivers}
		Leveraging the vocabulary, our system generates a vocabulary framework 
		 to aid the device developer~(step~\circled{8} in Figure~\ref{fig:devcycle}). The vocabulary framework contains {\em interfaces} 
		and {\em concrete classes} corresponding to resources defined in the vocabulary.  The concrete classes 
		contain concrete methods for interacting with other software components and platform-specific 
		device drivers. The interfaces are implemented by the device developer to write platform-specific 
		device drivers.

\end{itemize}

\subsubsection{Linking}

The linker combines and packs code generated by various stages into packages that can be deployed on devices.  
It merges generated architecture framework, application logic, mapping files, device drivers, 
and vocabulary framework~(step~\circled{9} in Figure~\ref{fig:devcycle}). This  stage supports the application 
deployment phase by producing device-specific code to result in a distributed software system collaboratively 
hosted by individual devices, thus providing automation 
at  the deployment phase\footnote{We assume that a 
middleware is already installed on the deployed devices. The installed middleware enables 
inter-device communication among devices.}.

\subsubsection{Handling evolution}
Evolution is an important aspect in IoT application development where new resources 
and computational services are added, removed, or extended. 
To deal with these changes, our development framework 
separates IoT application development into different concerns and allows an 
iterative development~\cite{Sommerville10} for these concerns.

This next section provides the details of our approach including three modeling languages~(SVL, SAL, and SDL),  
programming frameworks  to aid stakeholders, and an approach 
for handling evolution. This section refers to the building automation domain discussed 
in Section~\ref{sec:appexample} for describing examples. 
 

\subsection{Specifying domain concern with the SVL}\label{sec:specdomain}

The domain concern describes an application domain of an IoT application. The domain expert specifies 
it using SVL. A vocabulary includes specification of resources that 
are responsible for interacting with entities of interest, including sensors, 
actuators, storages, and user interfaces. Moreover, it includes region definitions specific to the 
application domain. We now present SVL for describing the domain concern.

SVL is designed to enable the domain expert to describe a domain vocabulary domain. It
offers constructs to specify concepts that interact with entities of  interest. 
Figure~\ref{fig:classdiagram-svl}  illustrates  domain-specific concepts 
(defined in the conceptual model Figure~\ref{fig:conceptualmodel}) 
that can be specified using SVL. These concepts can be described as 
$ \mathrm{V} = (\mathrm{P}, \mathrm{D}, \mathrm{R})$.  $\mathrm{P}$ represents 
the set of regions, $\mathrm{D}$ represents the set of data structure, and $\mathrm{R}$ represents 
the set of resources. We describe these concepts in detail as follows:

\begin{figure}[!ht]
 \centering
  \includegraphics[width=0.5\textwidth]{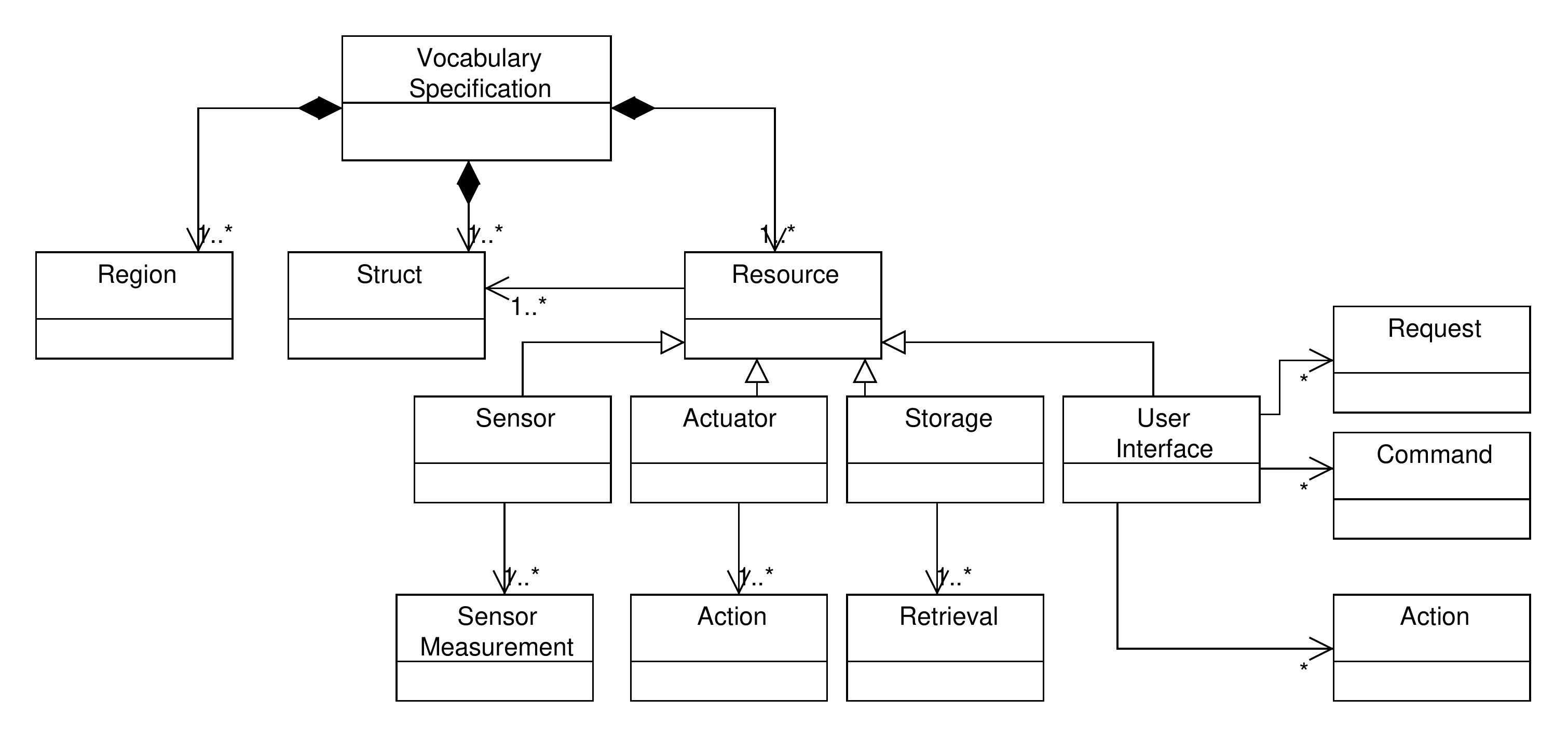}
	 \caption{Class diagram of domain-specific concepts}
	 \label{fig:classdiagram-svl}
 \end{figure}

     \fakeparagraph{{\em regions}~($\mathrm{P}$)} 
		  It represents the set of regions  that are used to specify locations of devices. 
			A region definition includes a region label and region type. For example, 
			the building automation is reasoned in terms of rooms and floors (considered as region labels), 
			while the transport domain is expressed in terms of highway sectors. Each room or floor in a 
			building may be annotated with an integer value (e.g. room:1 interprets as room number 1) considered 
			as region type.   This construct is declared using the  \texttt{regions} keyword. 
			Listing~\ref{vocabDSL} (lines~\ref{vocab:regions-start}-\ref{vocab:regions-end}) 
			shows region definitions for the building automation domain.
			
		  \fakeparagraph{{\em data structures}~($\mathrm{D}$)} Each resource is characterized by 
     types of information it generates or consumes.  A set of  information is defined using the \texttt{structs} 
		 keyword~(Listing~\ref{vocabDSL}, line~\ref{vocab:structs-start}). 	
		 For instance, a temperature sensor may generate  a	temperature value and 
		 unit of measurement~(e.g., Celsius or Fahrenheit). This information is defined 
		 as \texttt{TempStruct} and its two fields~(Listing~\ref{vocabDSL}, lines~\ref{vocab:tempstruct-start}-\ref{vocab:structs-end}).
		
	\fakeparagraph{{\em resources}~($\mathrm{R}$)} 
		It defines resources that might be attached with devices, including sensors, actuators, storages, or user interfaces. 
		It is defined as  $\mathrm{R} = (\mathrm{R}_{sensor}, \mathrm{R}_{actuator}, \mathrm{R}_{storage}, \mathrm{R}_{ui})$. 
		$\mathrm{R}_{sensor}$ represents a set of sensors,	$\mathrm{R}_{actuator}$ represents a set of actuators, 
		$\mathrm{R}_{storage}$ represents a set of storages, and $R_{ui}$ represents a set of user interfaces. 
		We describe them in detail as follows: 
			
		\begin{itemize}
			\item \textbf{{\em sensors}~($\mathrm{R}_{sensor}$)}: It defines a set of various types of 
			   sensors~(e.g., temperature sensor, smoke detector). A set of sensors is declared using the \texttt{sensors} keyword~(Listing~\ref{vocabDSL}, 
				 line~\ref{vocab:sensors-start}). $\mathrm{S}_{generate}$ is a set of sensor measurements produced  by  $\mathrm{R}_{sensor}$.
				 Each sensor~($\mathrm{S} \in \mathrm{R}_{sensor}$) produces 
				 one or more sensor measurements~($op \in \mathrm{S}_{generate}$) along with the data-types  specified in the data 
				 structure~($\mathrm{D}$). A sensor measurement of each sensor is declared using the \texttt{generate} keyword~(Listing~\ref{vocabDSL}, 
				 line~\ref{vocab:sensors-end}). For instance, a temperature sensor generates a temperature 
				 measurement of \texttt{Tempstruct} type~(lines~\ref{vocab:tempsensor-end}-\ref{vocab:sensors-end}) 
				 defined in data structures~(lines~\ref{vocab:tempstruct-start}-\ref{vocab:structs-end}). 
	
			\item \textbf{{\em actuators}~($\mathrm{R}_{actuator}$)}:  It defines a set of various types of actuator\footnote{				 Since a deployment infrastructure may be shared among a number of different IoT applications and users, it is 
			likely that these applications may have actuation conflicts. This work assumes 
			actuators are pre-configured which can resolve actuation conflicts.}~(e.g., heater, 
			  alarm). A set of actuators is declared using the \texttt{actuators} 
				keyword~(Listing~\ref{vocabDSL}, line~\ref{vocab:actuators-start}). 
				$\mathrm{A}_{action}$ is a set of actions performed by $\mathrm{R}_{actuator}$.
				Each actuator~($\mathrm{A} \in \mathrm{R}_{actuator}$) 
				has one or more actions~($a \in \mathrm{A}_{action}$) that is declared using the \texttt{action} keyword. 
				An action of an actuator may take inputs specified as parameters of an action (Listing~\ref{vocabDSL}, line~\ref{vocab:actuators-end}).	
				For instance, a heater may has two actions. One is to switch off the heater and second is to set the heater according to 
				a user's temperature preference illustrated in Listing~\ref{vocabDSL}, lines~\ref{vocab:heater}-\ref{vocab:actuators-end}.
				The \texttt{SetTemp} action takes a user's temperature preference shown in 
				line~\ref{vocab:actuators-end}. 	
				
					\item \textbf{{\em storages}~($\mathrm{R}_{storage}$)}:	It defines a set of storages\footnote{Even though IoT applications may include rich diverse set of storages available today 
on the Internet ({\em e.g.}, RDBMs and noSQL   databases, using content that is both user 
generated such as photos as well as machine generated such as sensor data), we restrict 
our work to key-value data storage services.}~(e.g., user's profile storage) that 
		   might be attached to a device. A set of storages is declared using the \texttt{storages} 
				keyword~(Listing~\ref{vocabDSL}, line~\ref{vocab:storages-start}). $\mathrm{ST}_{generate}$ represents 
				a set of retrievals of  $\mathrm{R}_{storage}$. A retrieval~($rq \in \mathrm{ST}_{generate}$)  
				from  the storage~($\mathrm{ST} \in \mathrm{R}_{storage}$) requires a parameter.  Such a parameter 
				is specified using the  \texttt{accessed-by} keyword~(Listing~\ref{vocabDSL}, line~\ref{vocab:storages-end}). 
				For instance,  a user's profile is accessed from profile storage by his unique badge identification 
				illustrated in Listing~\ref{vocabDSL}, lines~\ref{vocab:profile-start}-\ref{vocab:storages-end}.

				\item  \textbf{{\em user interfaces}~($\mathrm{R}_{ui}$)}: 
		       It defines a set of tasks (e.g., controlling a heater, receiving notification from a fire alarm, 
					  or requesting preference information from a database server) available to users to interact 
						with other entities. A set of user interfaces is declared using the \texttt{user interfaces} 
						keyword~(Listing~\ref{vocabDSL}, line~\ref{vocab:ui-start}). The user interface provides
						the following tasks:
						
						\begin{itemize}
					\item \textbf{\emph{command}~($\mathrm{U}_{command}$)}:			
						It is a set of commands available to users to control actuators, 
						represented as $\mathrm{U}_{command}$. 		
					 A user can control an actuator by triggering a command~(e.g., switch off the heater) declared using 
					  the  \texttt{command} keyword~(Listing~\ref{vocabDSL}, line~\ref{vocab:command-off}). 	
								
            \item \textbf{\emph{action}~($\mathrm{U}_{action}$)}: 
					 It is a set of actions that can be invoked by other entities to notify users,  
					 represented as $\mathrm{U}_{action}$. The other resources may notify a user~(e.g., notify the  current temperature) 
					 by invoking an action provided by the user interface. The notification task is declared using the \texttt{action} 
					 keyword~(Listing~\ref{vocabDSL}, line~\ref{vocab:displaydata}).
					
					\item \textbf{\emph{request}~($\mathrm{U}_{request}$)}: 
					 It is a set of request though which a user can request other resources for data, represented 
					 as $\mathrm{U}_{request}$. A user can retrieve data by requesting a resource~(e.g., retrieve my temperature preference). 
					 This is declared using the \texttt{request} keyword~(Listing~\ref{vocabDSL}, line~\ref{vocab:profile-ui}). 
		       \end{itemize}
		\end{itemize}

\lstset{emph={sensors, regions, string, generate, actuators, structs, double, structs, action, resources, long, storages, accessed,		command, with, user, interfaces, request, integer, -, by}, emphstyle={\color{blue}\bfseries\emph}, caption={Code snippet of the building automation domain using SVL. Keywords are printed in {\color{blue} \texttt{\textbf{blue}}}.}, escapechar=\#, 
 label=vocabDSL}	

\begin{lstlisting}
regions:  #\label{vocab:regions-start}#
    Building: integer;
    Floor: integer;
    Room: integer;#\label{vocab:regions-end}#
structs:                  #\label{vocab:structs-start}#
    BadgeDetectedStruct    #\label{vocab:badgedetectedstructs-start}#
        badgeID: string;
        timeStamp: long;   #\label{vocab:badgedetectedstructs-end}#
    TempStruct            #\label{vocab:tempstruct-start}#
        tempValue: double;
        unitOfMeasurement: string;  #\label{vocab:structs-end}#
resources:   
    sensors:                        #\label{vocab:sensors-start}#
        BadgeReader                 #\label{vocab:badgereader-start}# 
            generate badgeDetected: BadgeDetectedStruct;  #\label{vocab:badgereader-end}# 
        TemperatureSensor               #\label{vocab:tempsensor-end}#
            generate tempMeasurement: TempStruct;  #\label{vocab:sensors-end}#
    actuators:                   #\label{vocab:actuators-start}#
        Heater                 #\label{vocab:heater}#
            action Off();      #\label{vocab:action-end}#
            action SetTemp(setTemp: TempStruct);   #\label{vocab:actuators-end}#
    storages:       #\label{vocab:storages-start}#
        ProfileDB   #\label{vocab:profile-start}#
            generate  profile: TempStruct accessed-by badgeID: string;  #\label{vocab:storages-end}#
    userinterfaces:   #\label{vocab:ui-start}#
        EndUserGUI   #\label{vocab:endusergui-start}#
            command Off();  #\label{vocab:command-off}#
            action DisplayData(displayTemp: TempStruct);    #\label{vocab:displaydata}#     
            request profile(badgeID);    #\label{vocab:profile-ui}# 
\end{lstlisting}

The regions~($\mathrm{P}$), data structures~($\mathrm{D}$), and resources~($\mathrm{R}$) defined 
using SVL in the vocabulary are used to customize the grammar of SAL, 
and can be exploited by tools to provide support such as code completion to the software designer, discussed
next.

\subsection{Specifying functional concern}\label{sec:specfunc} 
This concern describes computational services and 
how they interact with each other to describe functionality of an application. 
We describe the computational services and interactions among them using SAL 
(discussed in Section~\ref{sec:arch}). The development framework customizes the SAL grammar  
to make domain-specific knowledge defined in the vocabulary available to the software designer 
and use it to generate an architecture framework. The application developer leverages this generated framework 
and implements the application logic on top of it~(discussed in Section~\ref{sec:impl}).  

\subsubsection{Srijan architecture language (SAL)}\label{sec:arch}

Based on a vocabulary, the SAL grammar is customized to
enable the software designer to design an application. Specifically, sensors~($\mathrm{R}_{sensor}$), 
actuators~($\mathrm{R}_{actuator}$), storages~($\mathrm{R}_{storage}$), user interfaces~($\mathrm{R}_{ui}$), 
and regions~($\mathrm{P}$) defined in the vocabulary become possible set of values for certain attributes in SAL (see
underlined words in Listing~\ref{list:archdsl}). 

\begin{figure}[!ht]
 \centering
  \includegraphics[width=0.5\textwidth]{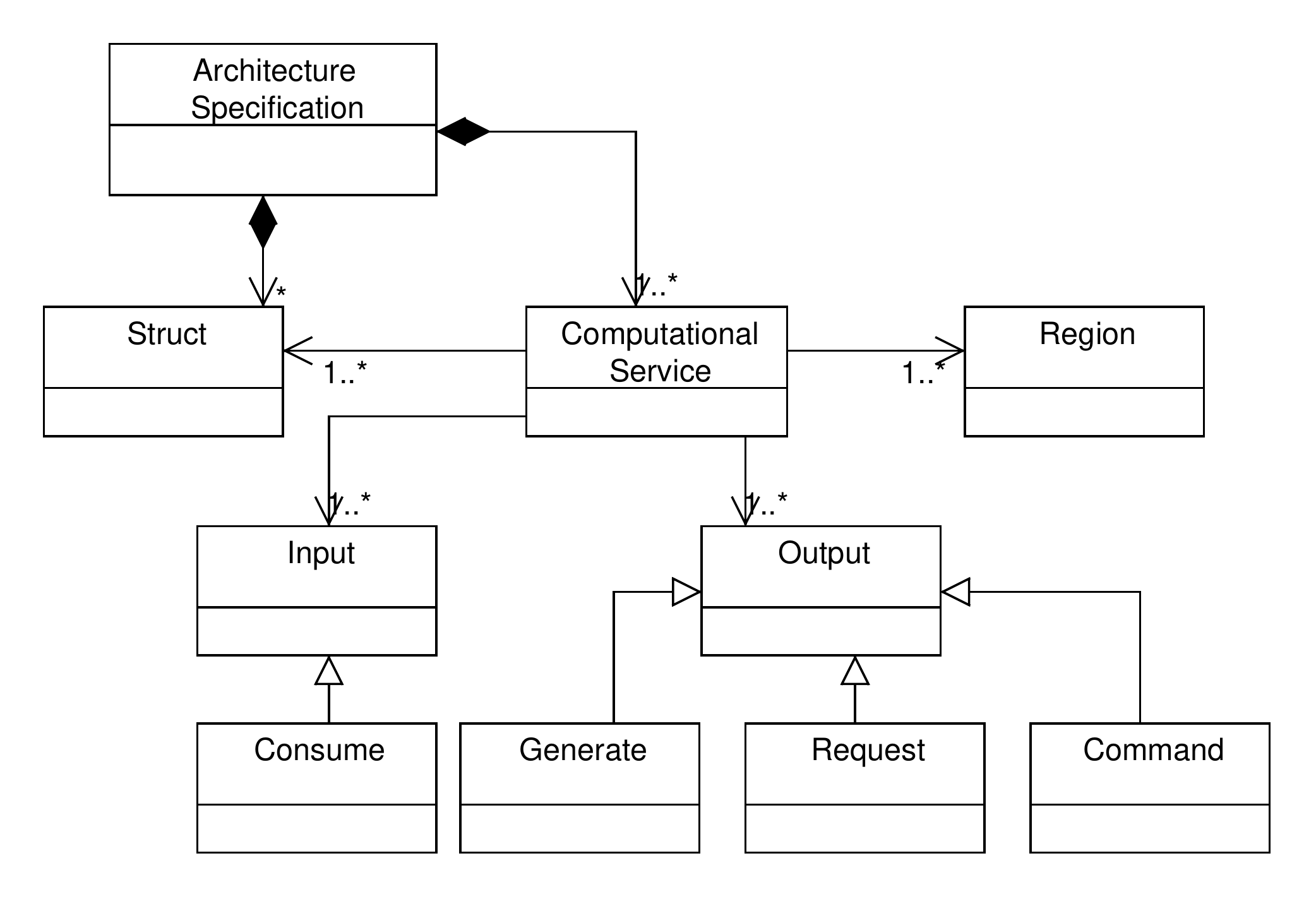}
	 \caption{Class diagram of functionality-specific concepts}
	 \label{fig:classdiagram-sal}
 \end{figure}

Figure~\ref{fig:classdiagram-sal}  illustrates concepts related-to 
a computational service that can be specified using SAL. It can be described as  
$\mathrm{A}_{v} = ( \mathrm{C} )$. $\mathrm{C}$ represents a set of computational services. 
It is described as $\mathrm{C} = (\mathrm{C}_{generate}, \mathrm{C}_{consume}, 
\mathrm{C}_{request}, \mathrm{C}_{command},  \mathrm{C}_{in-region}, \mathrm{C}_{hops})$. 
$\mathrm{C}_{generate}$ represents a set of outputs  produced by computational services. 
$\mathrm{C}_{consume}$  is a set of  inputs consumed by computational services. The inputs 
could be data produced by other computational services or sensors~($\mathrm{R}_{sensor}$). 
$\mathrm{C}_{request}$ represents a set of request by computational services to retrieve 
data from the storages~($\mathrm{R}_{storage}$). $\mathrm{C}_{command}$  represents a set of commands to 
invoke actuators~($\mathrm{R}_{actuator}$)  or user interfaces~($\mathrm{R}_{ui}$). 
$\mathrm{C}_{in-region}$ is a set of regions ($\mathrm{R}_{region}$) where 
computational services can be  placed. $\mathrm{C}_{hops}$ is a set 
of regions~($\mathrm{R}_{region}$) where computational services receive data.
In the following, we describe these concepts in detail.

\fakeparagraph{{\em consume}~($\mathrm{C}_{consume}$) and {\em generate}~($\mathrm{C}_{generate}$)}
These two concepts together define publish/subscribe interaction mode that provides subscribers 
with the ability to express their interest in an event, generated by a 
publisher, that  matches their registered interest.  A computational service represents the publish and subscribe 
using \emph{generate} and \emph{consume} concept respectively.  We describe these two concepts in details as follows:
\begin{itemize}
	\item \textbf{{\em consume}}:
		It represents a set of subscriptions~(or consumes) expressed by computational services to 
		get event notifications generated by sensors~($\mathrm{S}_{generate}$) defined in  
		the vocabulary specification  or other computational services~($\mathrm{C}_{generate}$) defined in 
		the architecture specification. Thus, $\mathrm{C}_{consume}$ can be $\mathrm{C}_{generate} \cup \mathrm{S}_{generate}$.  
		A consume~($ c \in \mathrm{C}_{consume}$) of a computational service is expressed using the \texttt{consume} keyword. The computational service 
		expresses its interest by an event name. For instance, a computational service \texttt{RoomAvgTemp}, 
		which calculates an average temperature of a room, subscribes its interest by expressing 
		event name \texttt{tempMeasurement} illustrated  in 
		Listing~\ref{list:archdsl}, line~\ref{arch-roomavgtemp-consume}.

\item \textbf{{\em generate}}: It represents a set of publications~(or generates) that 
	       are produced by computational services. A generate~($ g \in \mathrm{C}_{generate}$)
				 of a computational service is expressed using the \texttt{generate} keyword. 	
				 The computational service transforms data to be consumed by other computational services 
				 in accordance with the application needs. For instance, the computational service \texttt{RoomAvgTemp} 
				 consumes temperature measurements (i.e., \texttt{tempMeasurement}), calculates an average temperature of a room, and
				 generates \texttt{roomAvgTempMeasurement}~(Listing~\ref{list:archdsl}, lines~\ref{arch-roomavgtemp}-\ref{arch-roomavgtemp-consume}) 
				 that is used by \texttt{RoomController} service~(Listing~\ref{list:archdsl}, 
				 lines~\ref{arch-controller:regulatetemp-start}-\ref{arch-controller:regulatetemp}). 
\end{itemize}

\fakeparagraph{{\em request}~($\mathrm{C}_{request}$)} 
It is a set of requests, issued by computational services, 
to retrieve data from storages~($\mathrm{R}_{storage}$) defined in the vocabulary specification. 
A request is a one-to-one synchronous interaction with a return values. 
In order to fetch data, a requester sends a request message containing 
an access parameter to a responder. The responder receives and processes 
the request message, ultimately returns an appropriate message as a response. 
An access~($rq \in \mathrm{C}_{request}$) of the computational service 
is specified using \texttt{request} keyword. For instance, a computational 
service \texttt{Proximity}~(Listing~\ref{list:archdsl}, line~\ref{arch-proximity:proximity-request}), 
which wants to access user's profile data, sends a request message containing profile 
information as an access parameter to a storage \texttt{ProfileDB}~(Listing~\ref{vocabDSL}, line~\ref{vocab:storages-end}).  

		 
\fakeparagraph{{\em command}~($\mathrm{C}_{command}$)} 
It is a set of commands,  issued by a computational service to trigger actions provided by 
actuators~($\mathrm{R}_{actuator}$) or user interfaces~($\mathrm{R}_{ui}$).
So, it can be a subset of $\mathrm{A}_{action} \cup~\mathrm{U}_{action}$.
The software designer can pass arguments to a command depend on action signature provided 
by actuators or user interfaces.  Moreover, he specifies a scope of command, which specifies  
a region where commands are issued. A command is specified 
using the \texttt{command} keyword.  An example of command invocation is given in line~\ref{arch-proximity:proximity-command} 
of Listing~\ref{list:archdsl}.  The room controller service~(i.e., \texttt{roomController}), which regulates 
temperature, issues a \texttt{SetTemp} command with a preferred temperature 
as an argument~(i.e., \texttt{settemp}) to heaters~(Listing~\ref{vocabDSL}, line~\ref{vocab:actuators-end}).

\begin{figure}[!ht]
 \centering
  \includegraphics[width=0.4\textwidth]{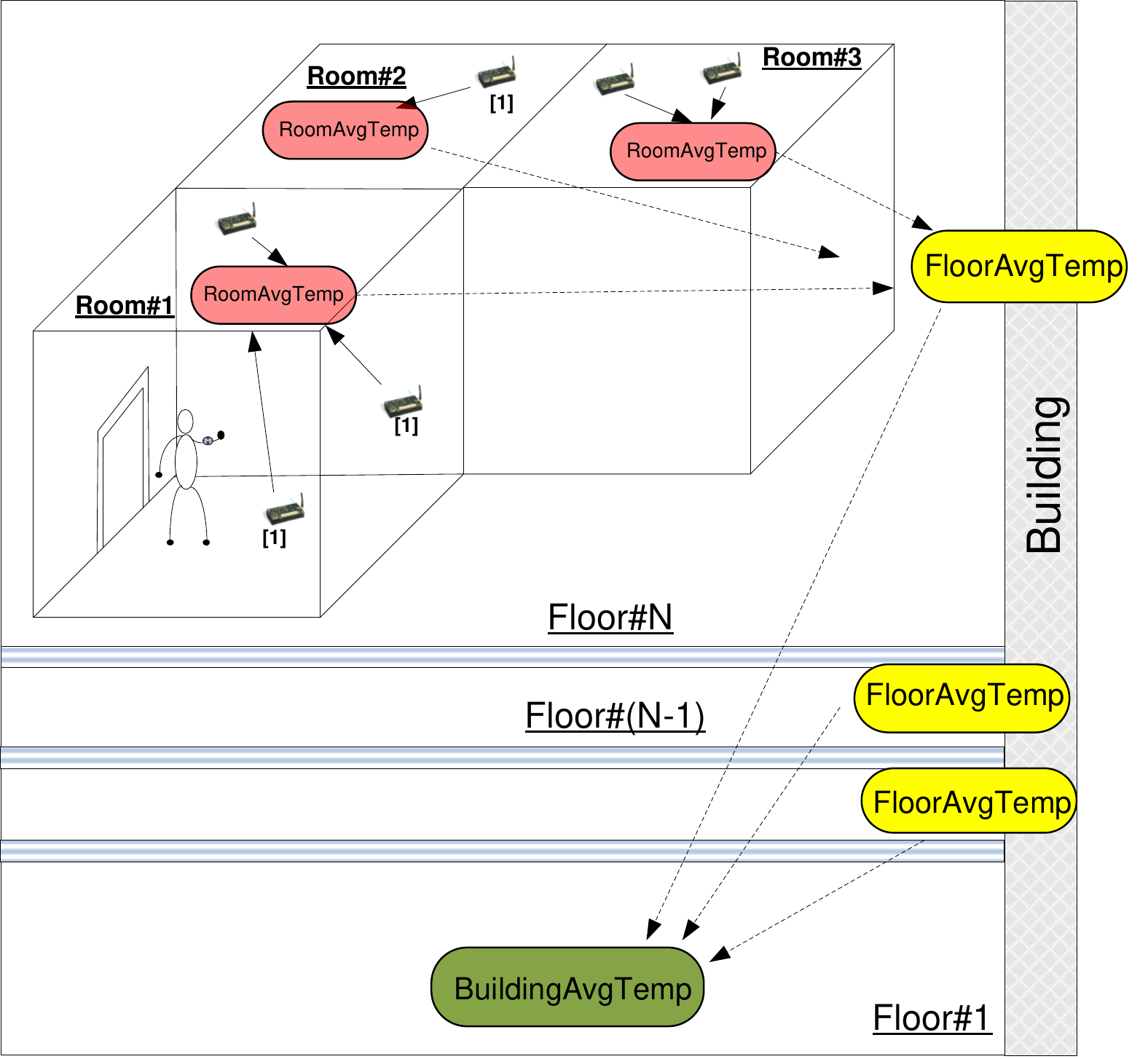}
	 \caption{Clustering in the  smart building application.  The device with temperature 
	    sensor is numbered as [1].}
	 \label{fig:clustering}
 \end{figure}

\fakeparagraph{{\em in-region}~($\mathrm{C}_{in-region}$) and \textbf{{\em hops}~($\mathrm{C}_{hops}$)}} 
To facilitate the scalable operations within an IoT application, devices should be grouped 
to form a cluster based on their spatial relationship~\cite{SINAmiddlewareshen2001}~(e.g.,``devices are in room\#1''). 
The grouping could be recursively applied to form a hierarchy of clusters. Within a cluster, a computational 
service is placed to receive and process data from its cluster of interest. Figure~\ref{fig:clustering} 
shows this concept for more clarity. The temperature data is first routed to a local average temperature 
service~(i.e., \texttt{RoomAvgTemp}), deployed in per room, then later per floor (i.e., \texttt{FloorAvgTemp}), and then 
ultimately routed to building average temperature service (i.e., \texttt{BuildingAvgTemp}).  
 
SAL offers {\em scope} constructs to define both the service 
placement~($\mathrm{C}_{in-region}$) and its data interest~($\mathrm{C}_{hops}$). The service placement 
(defined using the \texttt{in-region} keyword) is used to govern a placement of computational service 
in a cluster. The service placement can be in regions defined in a vocabulary specification.
So, it is a subset of $\mathrm{P}$.

The data interest of a computational service is used to define a cluster from which 
the computational service wants to receive data. The data interest can be in regions 
defined in the vocabulary specification. So, it is a subset of~$\mathrm{P}$.
It is defined  using the \texttt{hops} keyword. The  syntax of this keyword is \texttt{hops:radius:unit 
of radius}. Radius  is an integer value. The unit of radius is a cluster value. 
For example,  if  a computational service \texttt{FloorAvgTemp}  deployed  on floor number 12 has 
a data interest  \texttt{hops:i:Floor}, then it wants data from  all floors starting from \texttt{12}-th floor  
to \texttt{(12+i)}-th floor, and  all floors starting from  \texttt{12}-th floor  to \texttt{(12-i)}-th floor .

\begin{figure}[!ht]
	\centering
        \includegraphics[width=0.5\textwidth]{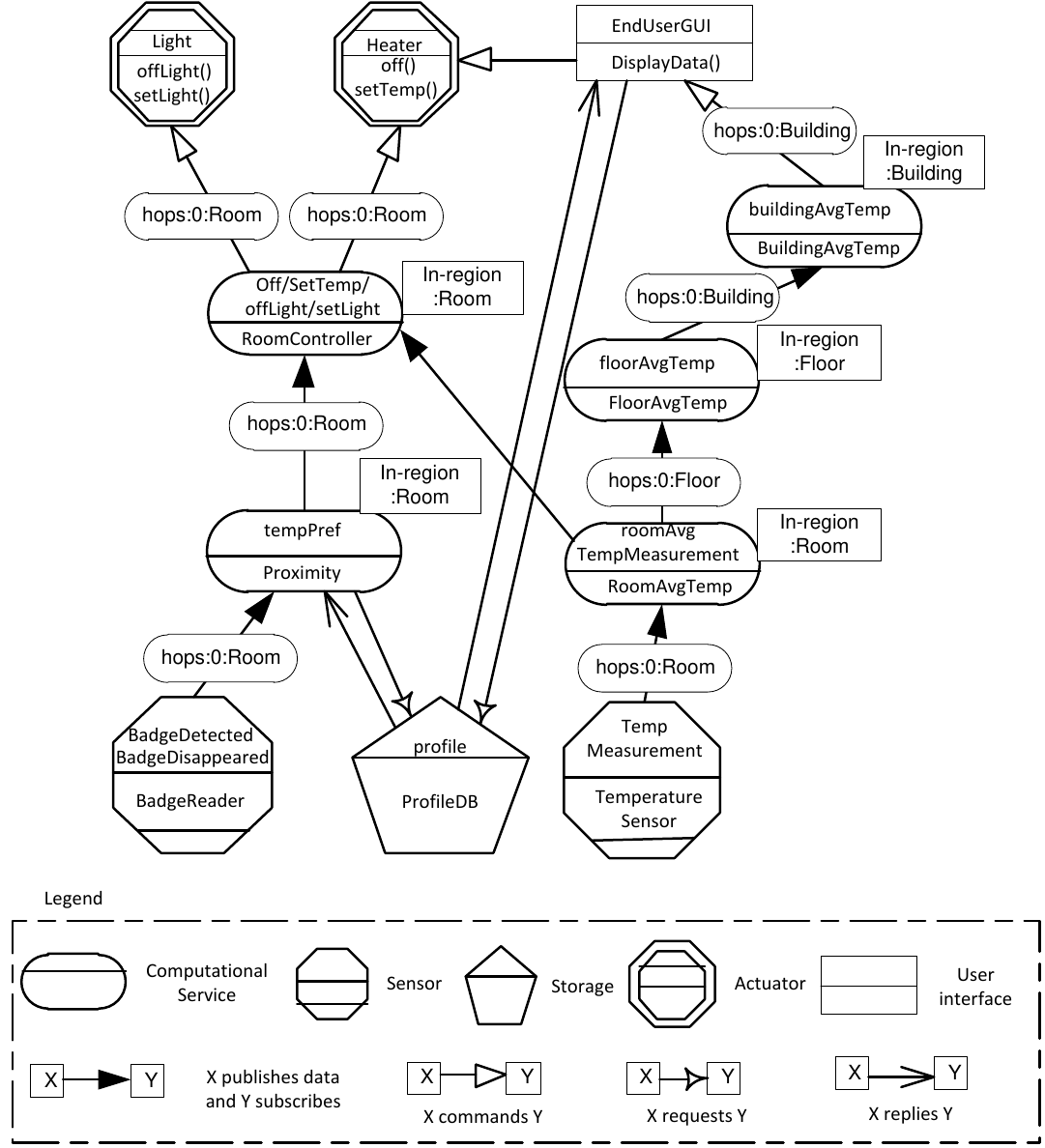}
	\caption{Architecture of the smart building application, similar to work in~\cite{towardscassou2011}.}
	\label{fig:dataflowOfScenario}
\end{figure}

Figure~\ref{fig:dataflowOfScenario} shows the architecture 
of the smart building application. Computational services are fueled
by sensing components. They process inputs data and take appropriate
decisions by triggering actuators. We illustrate SAL by examining a code
snippet in Listing~\ref{list:archdsl}, which describes a part of 
Figure~\ref{fig:dataflowOfScenario}. This
code snippet revolves around the actions of the \texttt{Proximity}
service (Listing~\ref{list:archdsl}, lines
\ref{arch-proximity:proximity-start}-\ref{arch-proximity:proximity-end}),
which coordinates events from the \texttt{BadgeReader} with the
content of \texttt{ProfileDB} storage service.  To do so,
the \texttt{Proximity} composes information from two sources, one for
badge events (i.e., badge detection), and one for requesting the
user's temperature profile from \texttt{ProfileDB}, expressed
using the \texttt{request} keyword (Listing~\ref{list:archdsl},
line~\ref{arch-proximity:proximity-request}). Input data is declared
using the \texttt{consume} keyword that takes source name and data
interest of  a computational service from logical region (Listing~\ref{list:archdsl},
line~\ref{arch-proximity:proximity-badgedetected-consume}).  The
declaration of \texttt{hops:0:room} indicates that the computational service is
interested in consuming badge events of the current room. The
\texttt{Proximity} service is in charge of managing badge events of
room.  Therefore, we need \texttt{Proximity} service to be
partitioned per room using \texttt{in-region:room}
(Listing~\ref{list:archdsl}, line~\ref{arch-proximity:proximity-end}).
The outputs of the \texttt{Proximity} and \texttt{RoomAvgTemp} are
consumed by the \texttt{RoomController} service~(Listing~\ref{list:archdsl},
lines~\ref{arch-controller:regulatetemp-start}-\ref{arch-controller:regulatetemp-end}).
This service is responsible for taking decisions that are carried
out by invoking commands declared using the
\texttt{command} keyword (Listing~\ref{list:archdsl},
line~\ref{arch-proximity:proximity-command}).

\lstset{emph={sensors, regions, string, generate, actuators, structs, 
							double, structs, action, abilities, controller, storages, 
							softwarecomponents, computationalService, command, from, integer, to, 
							consume, hops, devices, type, long, request, region, in, -
            }, 
						emphstyle={\color{blue}\bfseries\emph}
}

\lstset{emph={[2] Floor, Building, Room, tempMeasurement, badgeDetected, 
						badgeDisappeared, profile, Off, SetTemp, Display, setTemp, badgeID
						}, 
						emphstyle={[2] \color{black}\underbar} 
} 
\lstset{caption={A code snippet of the architecture specification for the smart building application using SAL. 
								The language keywords are printed in {\color{blue} {\texttt{\textbf{blue}}}}, while the keywords derived 
								from vocabulary are printed \underline{underlined}. }, 
								escapechar=\#, 	label=list:archdsl
}
\begin{lstlisting}
computationalServices: 
		Proximity                                   #\label{arch-proximity:proximity-start}#
			generate tempPref: UserTempPrefStruct;
			consume badgeDetected from hops:0:Room; #\label{arch-proximity:proximity-badgedetected-consume}#
			request profile(badgeID);                        #\label{arch-proximity:proximity-request}#
			in-region:Room;                         #\label{arch-proximity:proximity-end}#
		RoomAvgTemp#\label{arch-roomavgtemp}#
			generate roomAvgTempMeasurement:TempStruct;  #\label{arch-roomavgtemp-generate}#
			consume tempMeasurement from hops:0:Room ;   #\label{arch-roomavgtemp-consume}#
			in-region:Room;
		RoomController                            #\label{arch-controller:regulatetemp-start}#
			consume roomAvgTempMeasurement from hops:0:Room; #\label{arch-controller:regulatetemp}#
			consume tempPref from hops:0:Room;
			command SetTemp(setTemp) to hops:0:Room;  #\label{arch-proximity:proximity-command}#
			in-region:Room;                         #\label{arch-controller:regulatetemp-end}#
\end{lstlisting}

\subsubsection{Implementing application logic}\label{sec:impl}
Leveraging the architecture specification, we
generate a framework to aid the application developer. 
The generated framework contains abstract classes 
corresponding to the architecture specification. The abstract classes include two 
types of methods: (1)~\emph{concrete methods} to interact  with other components 
transparently through the middleware and  (2)~\emph{abstract methods} that allow the application developer to program 
the application logic.  The application developer 
implements each abstract method of generated abstract class. 
The key advantage of this framework is that a framework structure remains uniform. 
Therefore, the application developer have to know only locations of abstract methods 
where they have to specify the application logic.

\begin{figure}[!ht]
\centering
\includegraphics[width=0.4\textwidth]{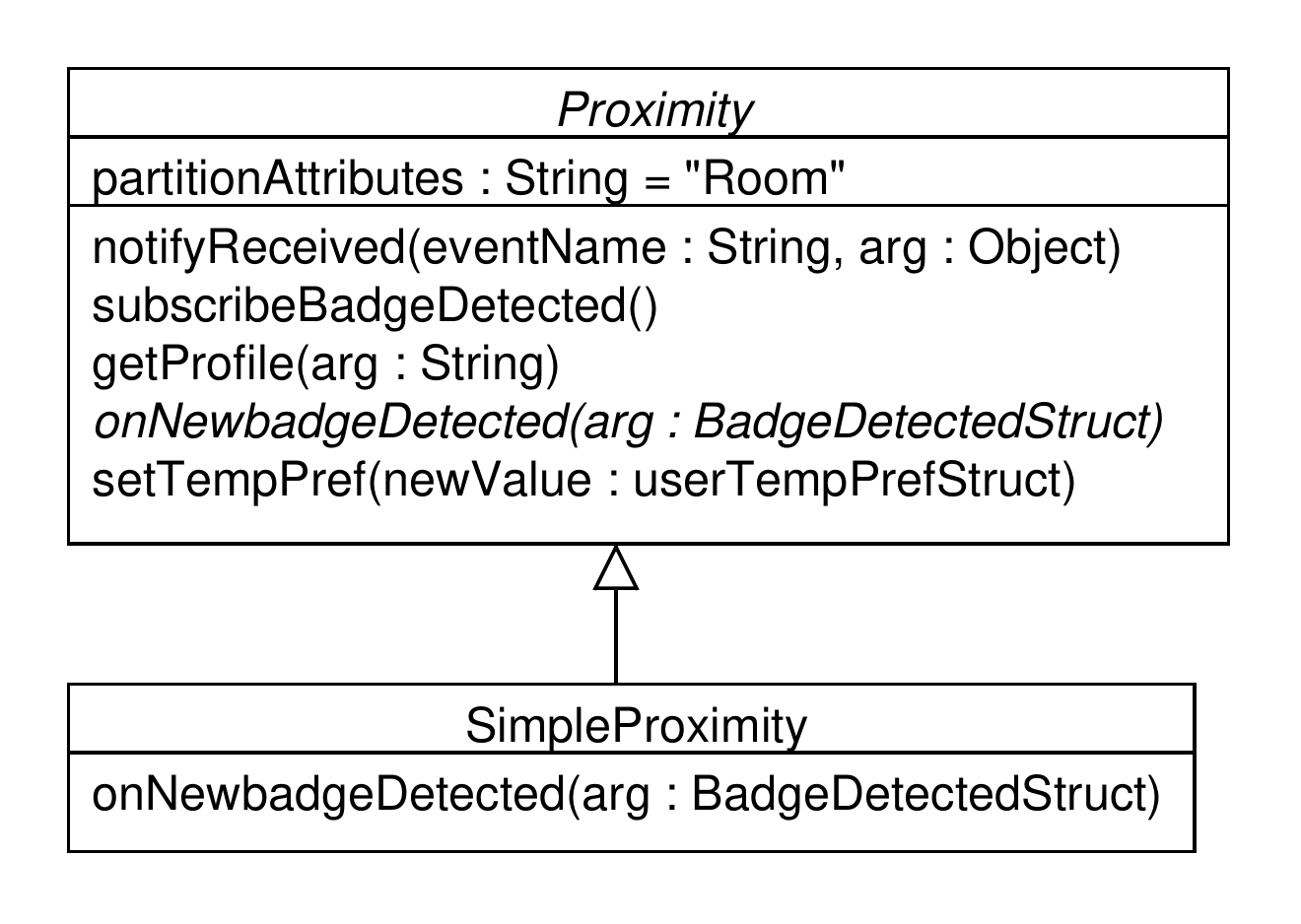}
\caption{Class diagram represents (1) the abstract class \texttt{Proximity} with its abstract method \texttt{onNewbadgeDetected()} 
illustrated in \textit{italicized} text, and  (2) the concrete implementation 
of \texttt{onNewbadgeDetected()} method is the \texttt{SimpleProximity} class.}
\label{fig:applicationlogic}
\end{figure}

\fakeparagraph{Abstract methods} For each input declared by a
computational service, an abstract method is generated for receiving data. This
abstract method is then implemented by the application developer. The class diagram in Figure~\ref{fig:applicationlogic} illustrates 
this concept. This class diagram uses \textit{italicized} text for the \texttt{Proximity} 
class, which represents an abstract class, and \texttt{onNewbadgeDetected()} 
that represents  abstract method. Then, it is implemented 
in the \texttt{SimpleProximity} class.

Listing~\ref{proximity} and \ref{fakeproximity} show Java 
code corresponding to the class diagram illustrated in 
Figure~\ref{fig:applicationlogic}. From the {\tt badgeDetected} input of
the \texttt{Proximity} declaration in the architecture specification
(Listing~\ref{list:archdsl}, lines~\ref{arch-proximity:proximity-start}-\ref{arch-proximity:proximity-end}),
the \texttt{onNewbadgeDetected()} abstract method is generated
(Listing~\ref{proximity}, line~\ref{generated-proximity:abs-onNewbadgeDetected}).  This method
is implemented by the application developer. Listing~\ref{fakeproximity} illustrates the implementation of
\texttt{onNewbadgeDetected()}.  It updates a user's temperature
preference and sets it using \texttt{settempPref()} method.

\lstset{language=Java, caption={The Java abstract class
    \texttt{Proximity} generated from the
    declaration \texttt{Proximity} in the
    architecture specification.}, escapechar=\#, label=proximity}
\begin{lstlisting}
public abstract class  Proximity {
    private String partitionAttribute = "Room";  #\label{generated-proximity:abs-partitionAttribute}#
    public void notifyReceived(String eventName, Object arg) { #\label{generated-proximity:abs-notifyReceived-start}# 
        if (eventName.equals("badgeDetected")) {
            onNewbadgeDetected((BadgeDetectedStruct) arg);
        } 
    }   #\label{generated-proximity:abs-notifyReceived-end}# 
    public void subscribebadgeDetected() { #\label{generated-proximity:abs-subscribebadgeDetected-start}# 
	    Region regionInfo = getSubscriptionRequest(
		      partitionAttribute, getRegionLabels(), getRegionIDs());
        PubSubMiddleware.subscribe(this, "badgeDetected", regionInfo);	
    } #\label{generated-proximity:abs-subscribebadgeDetected-end}#     
    protected TempStruct getprofile(String arg) {  #\label{generated-proximity:abs-getprofile-start}# 
        return (TempStruct) PubSubMiddleware.sendCommand("getprofile", arg, myDeviceInfo);
	} #\label{generated-proximity:abs-getprofile-end}#
    protected abstract void onNewbadgeDetected(BadgeDetectedStruct arg);   #\label{generated-proximity:abs-onNewbadgeDetected}# 	
    protected void settempPref(UserTempPrefStruct newValue) {  #\label{generated-proximity:abs-settempPref-start}# 
        if (tempPref != newValue) {
            tempPref = newValue;
        PubSubMiddleware.publish("tempPref", newValue, myDeviceInfo);
		} #\label{generated-proximity:abs-settempPref-end}# 
	}  
}
\end{lstlisting}

\lstset{language=Java, caption={The concrete implementation of the Java
    abstract class \texttt{Proximity} from
    Listing~\ref{proximity}, written by the application developer.}, escapechar=\#, label=fakeproximity}
\begin{lstlisting}
public class SimpleProximity extends Proximity {
    public void onNewbadgeDetected(BadgeDetectedStruct arg) {		 #\label{generated-fakeproximity:subclass-onNewbadgeDetected-start}#       
        long timestamp = ((long) (System.currentTimeMillis())) * 1000000; 	
        UserTempPrefStruct userTempPref = new UserTempPrefStruct(
			arg.gettempValue(), arg.getunitOfMeasurement(), timestamp);             
		settempPref(userTempPref);
	}                                            #\label{generated-fakeproximity:subclass-onNewbadgeDetected-end}# 
}
\end{lstlisting}

\fakeparagraph{Concrete methods} The compilation of an architecture specification generates concrete methods
to interact with other software component transparently. 
The generated concrete methods has the following two advantages:

\begin{enumerate}
\item {\bf Abstracting heterogeneous interactions.} 
To abstract heterogeneous interactions among software components, a compiler 
generates concrete methods that takes care of heterogeneous 
interactions. 
For instance,  a computational service processes input data and produces refined data  
to its consumers. The input data is either notified by other component~(i.e., 
publish/subscribe) or  requested~(i.e., request/response) by the service 
itself. Then, outputs are published. The concrete methods for these 
interaction modes are generated in an architecture framework. The lines~\ref{arch-proximity:proximity-start}~to~\ref{arch-proximity:proximity-end} 
of~Listing~\ref{list:archdsl} illustrates these heterogeneous interactions. 
The \texttt{Proximity} service has two inputs: (1) It receives \texttt{badgeDetect} 
event~(Listing~\ref{list:archdsl}, line~\ref{arch-proximity:proximity-badgedetected-consume}).
Our framework generates the \texttt{subscribebadgeDetected()} method to subscribe \texttt{badgeDetected} 
event~(Listing~\ref{proximity}, lines~\ref{generated-proximity:abs-subscribebadgeDetected-start}-\ref{generated-proximity:abs-subscribebadgeDetected-end}).
Moreover, it generates the implementation of \texttt{notifyReceived()} method to receive the published events~(Listing~\ref{proximity}, lines~\ref{generated-proximity:abs-notifyReceived-start}-\ref{generated-proximity:abs-notifyReceived-end}).
(2) It requests \texttt{profile} data~(Listing~\ref{list:archdsl},
line~\ref{arch-proximity:proximity-request}). A \texttt{sendcommand()} method 
is generated to request data from other components~(Listing~\ref{proximity}, 
lines~\ref{generated-proximity:abs-getprofile-start}-\ref{generated-proximity:abs-getprofile-end}).

\item {\bf Abstracting large scale.}
To address the scalable operations, a computational service annotates 
(1) its inputs with data interest, and (2) its placement in the region. 
Service placement and data interest jointly define a scope of a computational service to gather data.  
A generated architecture framework contains code that defines both data 
interest and its placement. For example, to get 
the \texttt{badgeDetected} event notification from the \texttt{BadgeReader}~(Listing~\ref{list:archdsl}, 
line~\ref{arch-proximity:proximity-badgedetected-consume}), the \texttt{subscribebadgeDetected()} 
method (Listing~\ref{proximity}, lines~\ref{generated-proximity:abs-subscribebadgeDetected-start}-\ref{generated-proximity:abs-subscribebadgeDetected-end}) 
is generated in the \texttt{Proximity} class. This method defines the data interest of a service from where it receives 
data. The value of \texttt{partitionAttribute}~(Listing~\ref{proximity}, line~\ref{generated-proximity:abs-partitionAttribute}), 
which comes from the architecture specification~(Listing~\ref{list:archdsl}, line~\ref{arch-proximity:proximity-end}), 
defines the scope of  receiving data. The above constructs are empowered by our choice of middleware, which is a variation of the one presented in~\cite{scopemottola2007}, and enables delivery of data across logical scopes.
\end{enumerate}

\subsection{Specifying deployment concern}\label{sec:specdeply}
This concern describes information about a target deployment  
containing various attributes of devices (such as location, type, attached resources) and 
locations where computational services are executed in a deployment, described using SDL~(discussed in Section~\ref{sec:network}). 
In order to map computational services to devices, we present a mapping technique that produces a mapping 
from a set of computational services to a set of devices~(discussed in Section~\ref{sec:mapping}).

\subsubsection{Srijan deployment language (SDL)}\label{sec:network}

\begin{figure}[!ht]
 \centering
  \includegraphics[width=0.3\textwidth]{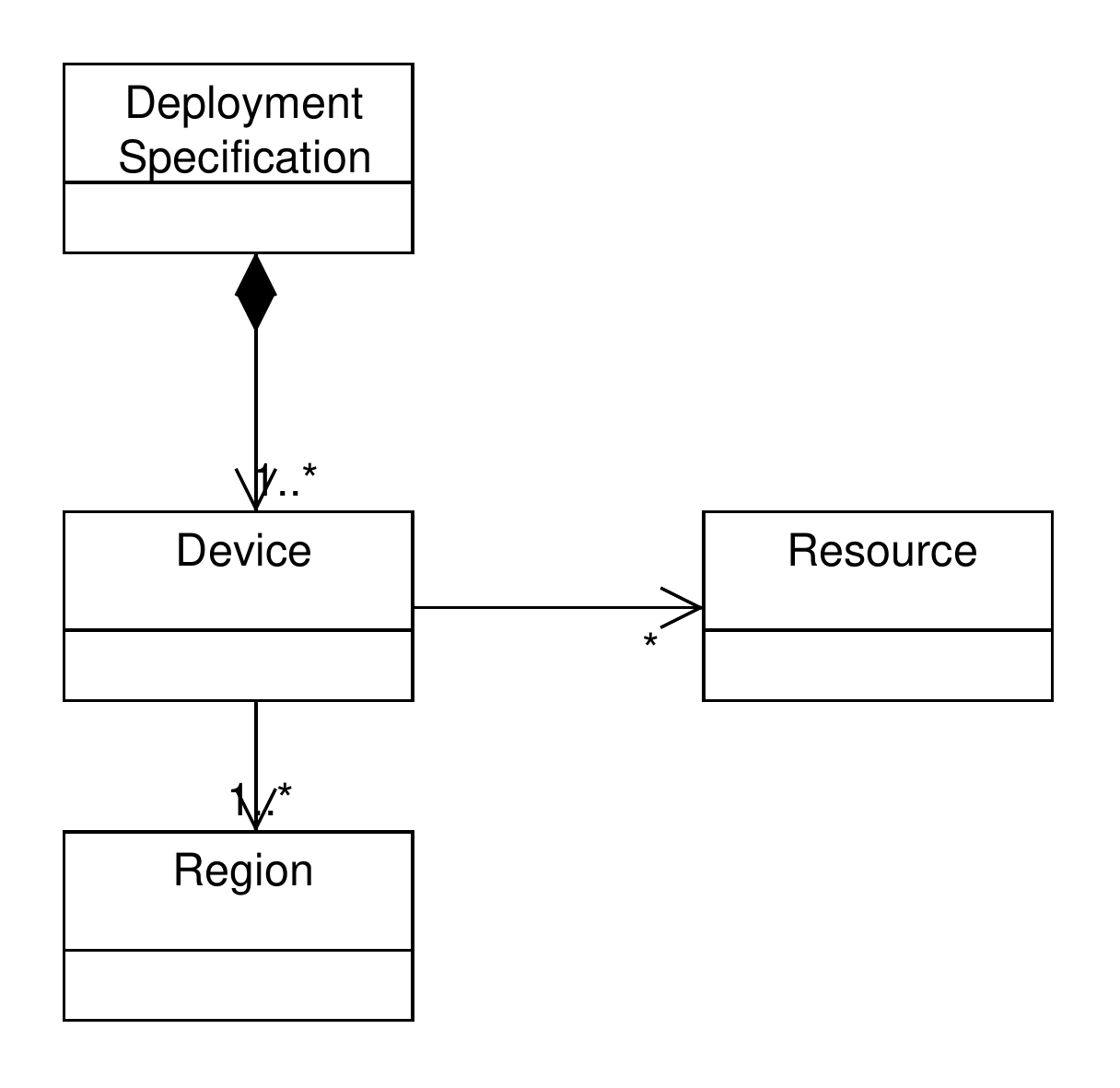}
	 \caption{Class diagram of deployment-specific concepts}
	 \label{fig:classdiagram-sdl}
 \end{figure}

Figure~\ref{fig:classdiagram-sdl}  illustrates  deployment-specific concepts 
(defined in the conceptual model Figure~\ref{fig:conceptualmodel}), specified using SDL.  
It includes device properties~(such as name, type), regions where devices are placed, and resources that are hosted by devices.
The resources~($\mathrm{R}$) and  regions~($\mathrm{P}$) defined  in a vocabulary become a
set of values  for certain attributes  in SDL~(see the underlined words  in Listing~\ref{list:networkdsl}). 
SDL can be described as  $\mathrm{T}_{v} = (D)$. $D$ represents  a set of  devices. 
A device~($d \in D$) can be defined as  ($\mathrm{D}_{region}, \mathrm{D}_{resource}, \mathrm{D}_{type}, 
\mathrm{D}_{mobile}$). $D_{region}$ represents a set of device placements in terms of regions 
defined  in a vocabulary. $D_{resource}$ is a subset of resources defined in a vocabulary. 
$D_{type}$ represents a set of device type~(e.g., JavaSE device, Android device) 
that is used to pick an appropriate device driver from a device driver repository. 
$D_{mobile}$ represents a set of two boolean values~(true or false).  The true value indicates 
a location of a device is not fixed, while the false value shows a fixed location.
Listing~\ref{list:networkdsl} illustrates a deployment specification 
of the smart building application. This snippet describes a device called \texttt{TemperatureMgmt-Device-1} 
with an attached \texttt{TemperatureSensor} and \texttt{Heater}, situated in \texttt{building}~15, \texttt{floor}~11, \texttt{room}~1, 
it is JavaSE enabled and non-mobile device.

Note that although individual listing of each device's attributes appears tedious, 
{\em i})~we envision that this information can be extracted from inventory logs 
that are maintained for devices purchased and installed in systems, and {\em ii)}~thanks 
to the separation between the deployment and functional concern in our approach, 
the same deployment specification can be re-used  across IoT applications 
of a given application domain.

\lstset{emph={type, sensors, regions,  String, generate, actuators, structs, 
double, structs, action, resources, controller, storages, 
softwarecomponents, computationalService, command, from, Integer, to, consume, 
hops, devices, long, request, region, in, mobile, false, true}, emphstyle={\color{blue}\bfseries\emph}%
}

\lstset{emph={[2] Floor, Building, Room, TemperatureSensor, 
BadgeReader, Monitor, Heater, ProfileDB}, emphstyle={ [2] \color{black}\underbar} } 

\lstset{caption={Code snippet of a deployment specification for the building automation domain using SDL. 
The language keywords are printed in {\color{blue} \texttt{\textbf{blue}}}, while the keywords derived 
from a vocabulary are printed \underline{underlined}.}, label=list:networkdsl}
\begin{lstlisting}
devices:
	TemperatureMgmt-Device-1:
		region:
			Building: 15 ;
			Floor: 11;
			Room: 1;
		resources: TemperatureSensor, Heater;
		type: JavaSE;
    mobile: false;
	...
\end{lstlisting}

\subsubsection{Mapping}\label{sec:mapping}

This section presents our mapping algorithm that decides devices 
for a placement of computational services. It takes inputs as (1) a list of devices~$D$ defined in 
a deployment specification~(see listing~\ref{list:networkdsl}) and 
(2) a list of  computational services~$C$ defined  in an architecture specification~(see listing~\ref{list:archdsl}).  
It produces a mapping of computational services to a set of devices. 

We presents the mapping algorithm (see Algorithm~\ref{mappingAlgo}) that comprises two steps.  The first 
step~(lines~\ref{mapping-first-ds-construct-start}-\ref{mapping-first-ds-construct-end})
constructs the two key-value data structures from a deployment specification. These two 
data structures are used in the second step. The second step~(lines~\ref{mapping-second-start}-\ref{mapping-second-end}) 
selects  devices randomly and allocates computational services to the selected devices\footnote{
A mapping algorithm cognizant of heterogeneity, associated with devices of a target deployment, 
is a part of our future work. See future work for detail.}. In order to give 
more clarity to readers, we describes these two steps in detail below.

The first step~(Algorithm~\ref{mappingAlgo}, lines~\ref{mapping-first-ds-construct-start}-\ref{mapping-first-ds-construct-end}) constructs  two key-value data structures $regionMap$ and $deviceListByRegionValue$ from $D$. 
The $regionMap$~(line~\ref{mapping-first-ds-construct-regionMap}) is a key-value data structure where $regionName$~(e.g., \texttt{Building}, \texttt{Floor}, 
\texttt{Room} in the listing~\ref{list:networkdsl}) is a key and $regionValue$~(e.g., \texttt{15}, \texttt{11}, \texttt{1} in the listing~\ref{list:networkdsl}) 
is  a value. The $deviceListByRegionValue$~(line~\ref{mapping-first-ds-construct-deviceListByRegionValue}) is a key-value data structure where 
$regionValue$ is a key  and  $device$~(e.g., \texttt{TemperatureMgmt-Device-1} in the listing~\ref{list:networkdsl}) is a value. Once, these two 
data structures are constructed, we use them for the second step~(lines~\ref{mapping-second-start}-\ref{mapping-second-end}).

The second step~(Algorithm~\ref{mappingAlgo}, lines~\ref{mapping-second-start}-\ref{mapping-second-end}) selects a device and allocates 
computational services to the selected device. To perform this task, the line~\ref{mapping-second-start} retrieves 
all keys~(in our example \texttt{Building}, \texttt{Floor}, \texttt{Room})  of $regionMap$ using   $getKeySet()$ function. 
For each computational service (e.g., \texttt{Proximity}, \texttt{RoomAvgTemp}, \texttt{RoomController} 
in listing~\ref{list:archdsl}), the selected key from the $regionMap$ is compared with a partition value of a 
computational component~(line~\ref{mapping-second-compare}). If the value match,  the next 
step~(lines~\ref{mapping-second-random-start}-\ref{mapping-second-random-end})  
selects a device randomly and allocates a computational service to the selected device.

\fakeparagraph{Computational complexity}
The first  step~(Algorithm~\ref{mappingAlgo}, lines~\ref{mapping-first-ds-construct-start}-\ref{mapping-first-ds-construct-end}) takes  $O(mr)$ times, where $m$ is a number of devices and $r$ is  a number of region pairs in each device specification. 
The second step~(Algorithm~\ref{mappingAlgo}, lines~\ref{mapping-second-start}-\ref{mapping-second-end}) 
takes  $O(nks)$ times, where $n$  is  a  number of region names~(e.g., building, floor, room for the building automation domain) 
defined in a vocabulary, $k$ is a number of  computational services defined in an architecture specification, 
and $s$ is a number of region values specified in a deployment specification. 
Thus, total computational complexity of the mapping algorithm is $ O(mr + nks)$.

\begin{algorithm}[h]
\caption{Mapping Algorithm}
\label{mappingAlgo}
\begin{algorithmic}[1]
\footnotesize\ttfamily
\REQUIRE  List $D$ of $m$ numbers of devices, List $C$ of $k$ numbers computational services  \label{mapping-input}
\ENSURE   List $mappingOutput$ of $m$ numbers that contains assignment of $C$ to $D$     \label{mapping-output}

\STATE Initialize $regionMap$ key-value pair data structure
\STATE Initialize $deviceListByRegionValue$ key-value pair data structure
\STATE Initialize $mappingOutput$ key-value pair data structure

\FORALL{$device$ in $D$}  \label{mapping-first-ds-construct-start}
\FORALL{ pairs ($regionName$, $regionValue$) in $device$ }   
	\STATE $regionMap[regionName]\leftarrow regionValue$
	\COMMENT{construct $regionMap$ with $regionName$ as key and assign $regionValue$ as Value}  \label{mapping-first-ds-construct-regionMap}
	\STATE $deviceListByRegionValue[regionValue]\leftarrow device$ \label{mapping-first-ds-construct-deviceListByRegionValue}
\ENDFOR
\ENDFOR    \label{mapping-first-ds-construct-end}
\FORALL{$regionName$ in $regionMap.getKeySet()$}\label{mapping-second-start}
	\FORALL {$computationalservice$ in $C$}   \label{mapping-second-each-component}
	\IF{$computationalservice.partitionValue()$ = $regionName$}\label{mapping-second-compare}
	  \FORALL{$regionValue$ in $regionMap.getValueSet(region$ $Name)$}\label{mapping-second-random-start}
	   \STATE $deviceList$ $\leftarrow deviceListByRegionValue.getValueSet$ $(regionValue)$
	   \STATE $selectedDevice \leftarrow selectRandomDeviceFromList(d$ $eviceList)$
	   \STATE $mappingOutput[selectedDevice] \leftarrow computational$ $service$
      \ENDFOR 	  \label{mapping-second-random-end}
    \ENDIF	
	\ENDFOR	
\ENDFOR    \label{mapping-second-end}
\RETURN $mappingOutput$
\end{algorithmic}
\end{algorithm}

\subsection{Specifying platform concern}\label{sec:specplat}
This concern describes software components  that act as a translator between 
a hardware device and an application. Because these components  are operating system-specific, 
the device developer implements them by hand.  
To aid the device developer, we generate a vocabulary framework to implement 
platform-specific device drivers. In the following section, we describe it in more detail.

\subsubsection{Implementing device drivers}\label{sec:driver}
Leveraging the vocabulary specification, our system generates a vocabulary framework to aid 
the device developer. The vocabulary framework contains 
\emph{concrete classes} and \emph{interfaces} corresponding to resources defined in 
a vocabulary. A concrete class contains concrete methods 
for interacting with other software components and platform-specific 
device drivers. The interfaces are implemented by the device developer 
to write platform-specific device drivers. In order to enable interactions 
between concrete class and platform-specific 
device driver, we adopt the factory design pattern~\cite{Gamma1995}.
This pattern provides an interface for a concrete class to obtain an 
instance of different platform-specific device driver implementations without having to 
know what implementation the concrete class obtains. 
Since the platform-specific device driver implementation can be updated without 
necessitating any changes in code of concrete class, the factory pattern 
has advantages of encapsulation and code reuse. We illustrate this concept 
in the following paragraph with a \texttt{BadgeReader} example.

\begin{figure*}[!ht]
\centering
\includegraphics[width=1.0\textwidth]{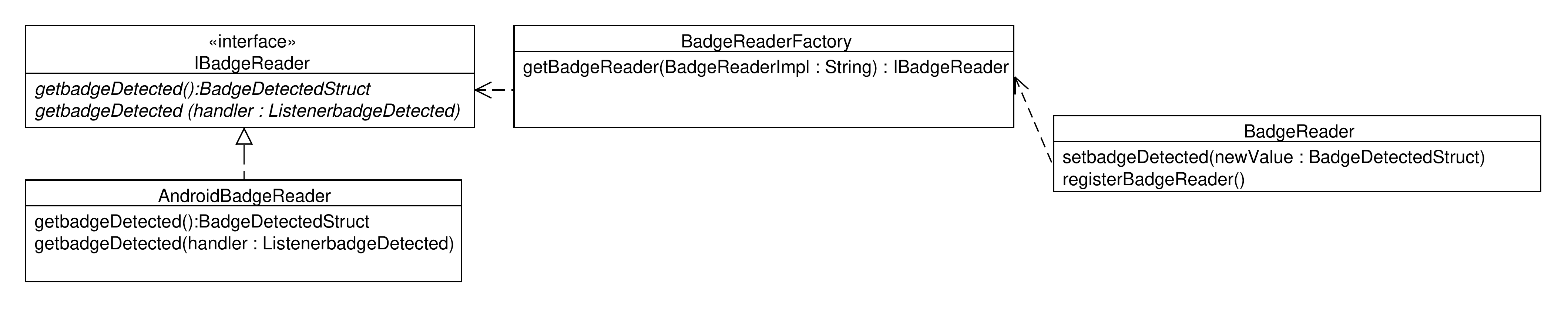}
\caption{Class diagram representing (1) the interface \texttt{IBadgeReader}  
and the implementation of two abstract methods in the \texttt{AndroidBadgeReader} class, 
(2) the concrete class \texttt{BadgeReader} that refers  the \texttt{AndroidBadgeReader}
through the \texttt{BadgeReaderFactory} factory class.}
\label{fig:classdiagramevolution}
\end{figure*}

The class diagram in Figure~\ref{fig:classdiagramevolution} illustrates the concrete 
class  \texttt{BadgeReader}, the interface~\texttt{IBadgeReader}, and  the associations 
between them through the factory class \texttt{BadgeReaderFactory}. The two abstract methods 
of the \texttt{IBadgeReader} interface~(Listing~\ref{ibadgereader}, lines~\ref{ibadgereader-start}-\ref{ibadgereader-end})  
are implemented in the \texttt{AndroidBadgeReader} class~(Listing~\ref{implbadgereader}, 
lines~\ref{implbadgereader-start}-\ref{implbadgereader-end}). 
The platform-specific implementation is accessed through the \texttt{BadgeReaderFactory} class~(Listing~\ref{badgereaderfactory},
lines~\ref{badgereaderfactory-start}-\ref{badgereaderfactory-end}). 
The \texttt{BadgeReaderFactory} class returns an instance of platform-specific implementations according to request 
by the concrete method \texttt{registerBadgeReader()} in the \texttt{BadgeReader} class~(Listing~\ref{badgereader},
lines~\ref{badgereader:registerBadgeReader-start}-\ref{badgereader:registerBadgeReader-end}).
In the following, we describe this class diagram with code snippet.

\fakeparagraph{Concrete class} For each resource declared in a vocabulary specification, a concrete class is 
generated. This class contains concrete methods for interacting with other components transparently~(similar to 
discussed in Section~\ref{sec:impl}) and for interacting with platform-specific implementations. 
For  example,  the \texttt{BadgeReader}~(Listing~\ref{badgereader}, lines~\ref{badgereader-start}-\ref{badgereader-end})  
class is generated from the \texttt{BadgeReader}  declaration~(Listing~\ref{vocabDSL},  
lines~\ref{vocab:badgereader-start}-\ref{vocab:badgereader-end}).
The generated class contains the \texttt{registerBadgeReader()} method~(Listing~\ref{badgereader},
lines~\ref{badgereader:registerBadgeReader-start}-\ref{badgereader:registerBadgeReader-end}).
This method first obtains a reference of one~(in our example Android) of platform-specific implementations, 
then uses that reference to create an object of that device-specific type~(Listing~\ref{badgereader},
line~\ref{badgereader:objBadgeReader-start}). This reference is used to disseminate badgedetect event~(Listing~\ref{badgereader},
lines~\ref{badgereader:ListenerbadgeDetected-start}-\ref{badgereader:ListenerbadgeDetected-end}).

\lstset{emph={BadgeReader}, 
						emphstyle={\color{black}}
}

\lstset{emph={[2] 
						}, 
						emphstyle={[2] \color{black}\underbar} 
}

\lstset{language=Java, caption={The Java 
    \texttt{BadgeReader} class generated from the \texttt{BadgeReader} declaration in
    the vocabulary specification.}, escapechar=\#, label=badgereader}
\begin{lstlisting}
public class BadgeReader {    #\label{badgereader-start}#
    protected void setbadgeDetected(BadgeDetectedStruct newValue) {
            ...
        PubSubMiddleware.publish("badgeDetected", newValue, DeviceInfo);		
    }	
    badgeDetected badgeDetectEvent = new badgeDetected() {  #\label{badgereader:ListenerbadgeDetected-start}#    
        public void onNewbadgeDetected(BadgeDetectSt resp) {
            BadgeDetectSt sBadgeDetectSt = new BadgeDetectSt(resp.getbadgeID(), resp.gettimeStamp());			
            publishbadgeDetectedEvent(sBadgeDetectSt);
        }
    };	 #\label{badgereader:ListenerbadgeDetected-end}#
    protected void registerBadgeReader(){   #\label{badgereader:registerBadgeReader-start}#    
       	IBadgeReader objBadgeReader = BadgeReaderFactory.getBadgeReader("Android");	 #\label{badgereader:objBadgeReader-start}#	
        objBadgeReader.getbadgeDetected(badgeDetectEvent);    #\label{badgereader:handlerBadgeReader-start}#
    }   #\label{badgereader:registerBadgeReader-end}#
}     #\label{badgereader-end}#
\end{lstlisting}

\fakeparagraph{Interfaces} 
For each resource declared in a vocabulary specification, interfaces are generated. Each interface
contains synchronous and asynchronous abstract methods corresponding to a resource declaration. These methods
are implemented by the device developer to write device-specific drivers. For example, our development system generates a 
vocabulary framework  that contains the interface \texttt{IBadgeReader}~(Listing~\ref{ibadgereader}, lines~\ref{ibadgereader-start}-\ref{ibadgereader-end})  
corresponding to the \texttt{BadgeReader}~(Listing~\ref{vocabDSL}, lines~\ref{vocab:badgereader-start}-\ref{vocab:sensors-end}) 
declaration in the vocabulary specification.  The device developer programs Android-specific implementations 
in the \texttt{AndroidBadgeReader} class by implementing the methods \texttt{getbadgeDetected()} and \texttt{getbadgeDetected(handler)} 
of the generated interface \texttt{IBaderReader}~(Listing~\ref{implbadgereader}, lines~\ref{implbadgereader-start}-\ref{implbadgereader-end}).

\lstset{language=Java, caption={The Java \texttt{BadgeReaderFactory} class.}, escapechar=\#, label=badgereaderfactory}
\begin{lstlisting}
public class BadgeReaderFactory {   #\label{badgereaderfactory-start}#  
    public static IBadgeReader getBadgeReader(String nameBadgeReader) {

     if(nameBadgeReader.equals("Android"))
         return new AndroidBadgeReader();    
            
      if (nameBadgeReader.equals("PC"))
	    return new PCBadgeReader();
    }
}  #\label{badgereaderfactory-end}#  
\end{lstlisting}

\lstset{language=Java, caption={The Java interface
    \texttt{IBadgeReader} generated from the  \texttt{BadgeReader} declaration in
    the vocabulary specification.}, escapechar=\#, label=ibadgereader}
\begin{lstlisting}
public interface IBadgeReader {       #\label{ibadgereader-start}#
	public  BadgeDetectedStruct getbadgeDetected();  #\label{ibadgereader-get}#       
	public  void getbadgeDetected(ListenerbadgeDetected handler);	#\label{ibadgereader-listener}#    
}  #\label{ibadgereader-end}#
\end{lstlisting}

\lstset{language=Java, caption={The device developer writes Android-specific 
        device driver of a badge reader by implementing the \texttt{IBadgeReader} interface.}, 
      escapechar=\#, label=implbadgereader}
\begin{lstlisting}
public class AndroidBadgeReader implements IBadgeReader { #\label{implbadgereader-start}#
	@Override
	public BadgeDetectedStruct getbadgeDetected() { #\label{implbadgereader:getbadgeDetected-start}#
        // The device developer implements platform-specific code here        
	}  #\label{implbadgereader:getbadgeDetected-end}#
	@Override
	public void getbadgeDetected(ListenerbadgeDetected handler) { #\label{implbadgereader:eventgetbadgeDetected-start}#
        // The device developer implements platform-specific code here        
	} #\label{implbadgereader:eventgetbadgeDetected-end}#
} #\label{implbadgereader-end}#
\end{lstlisting} 

\subsection{Handling evolution}\label{sec:handleevolution}

Evolution is an important aspect in IoT application development where resources 
and computational services are added, removed, or extended. 
To deal with these changes, we separate IoT application development into different concerns 
and allow an iterative development for these concerns. 
This iterative development requires only a change in evolved specification and reusing 
dependent specifications/implementation in compilation process, thus reducing effort 
to handle evolution, similar to the work in~\cite{towardscassou2011}. 

\begin{figure*}[!ht]
\centering
\includegraphics[width=0.72\textwidth]{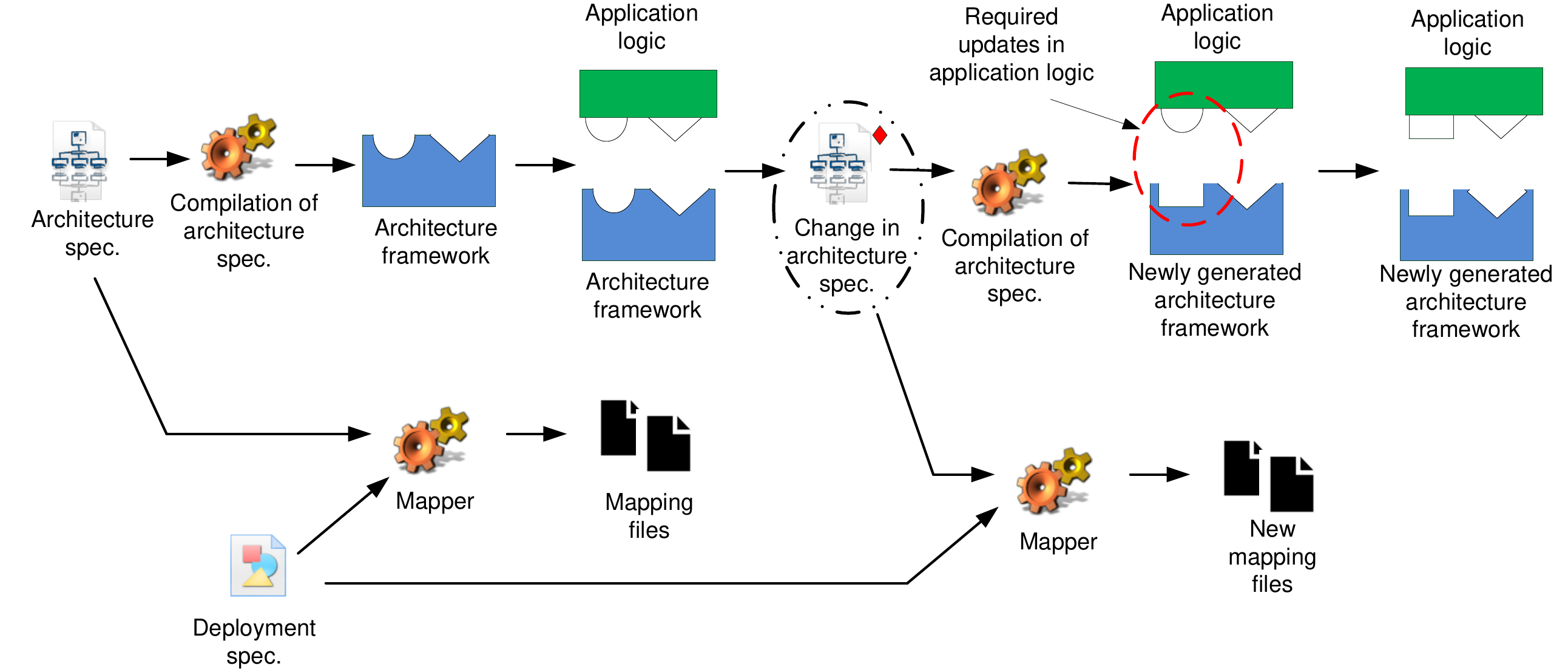}
\caption{Handling evolution in the functional concern}
\label{fig:evolutioninfunctionality}
\end{figure*}

Figure~\ref{fig:evolutioninfunctionality} illustrates evolution in the functional concern. 
It could be addition, removal, or extension of computational services.
A change in an architecture specification requires recompilation of it. The
recompilation generates a new architecture framework and preserves the previously 
written  application logic. This requires changes in the existing 
application logic implementations manually and re-compilation of the architecture specification to  
generate new mapping files that replaces old mapping files. We now review main evolution cases 
in each development concern and how our approach handles them.

\fakeparagraph{\emph{Changing functionality}} It refers to a change in behaviors 
of an application.  For example, while an application might be initially 
defined to switch on an air-conditioner when a temperature of room is greater 
than $30^{\circ} C$, a new functionality might be to open a window.  
This case requires to write a new architecture specification and application logic. 

\fakeparagraph{\emph{Adding a new computational service}}
It refers to the addition of a new computational service in an architecture 
specification. The application developer  implements  the application logic of the newly added 
computational services.

\fakeparagraph{\emph{Removing a computational service}}
It refers to the removal of an existing computation service from 
an architecture specification. The application developer has to manually 
remove application logic files of the removed computational service.

\fakeparagraph{\emph{Adding a new input source}}
A new input of a computational service, represented as \texttt{consume} keyword, 
can be added. The application developer implements a generated abstract method corresponding to 
a new input in application logic files. 

\fakeparagraph{\emph{Removing a input source}}
An input can be removed from a computational service.  In this case, the abstract method 
that implements the application logic becomes dead in application logic files. 
The IDE automatically reports errors. The application developer has to remove 
this dead abstract method manually.

\fakeparagraph{\emph{Removing an output  or command}}
An output~(\texttt{generate} keyword) or command~(\texttt{command} keyword) can be removed from 
an architecture specification. In this case code, which deals with output or command, becomes dead in 
application logic files. The application developer has to manually remove dead code.

%% file: component.tex
\section{Components of IoTSuite}\label{sec:components}

In the following, we demonstrate the application development process, mentioned in 
Section~\ref{sec:detail-ourapproach}, using IoTSuite\footnote{An open source version, targeting on 
Android- and JavaSE -enabled devices and MQTT middleware, is available on: 
\url{https://github.com/pankeshlinux/IoTSuite/wiki}} -- a suite of tools, which is composed 
of different components, mentioned below, at each phase of application development 
that stakeholders can use.

\fakeparagraph{\emph{Editor}} It helps stakeholders to write high-level specifications, including vocabulary, architecture, 
	    and deployment specification with the facilities of syntax coloring and syntax error reporting. We use  
Xtext\footnote{\url{http://www.eclipse.org/Xtext/}} for a full fledged editor support, similar 
to work in~\cite{bertran2012diasuite}. The Xtext  is a framework for a development of domain-specific 
languages, and provides an editor with syntax coloring by writing Xtext grammar.

\begin{figure*}[!ht]
	\centering \includegraphics[width=0.6\linewidth]{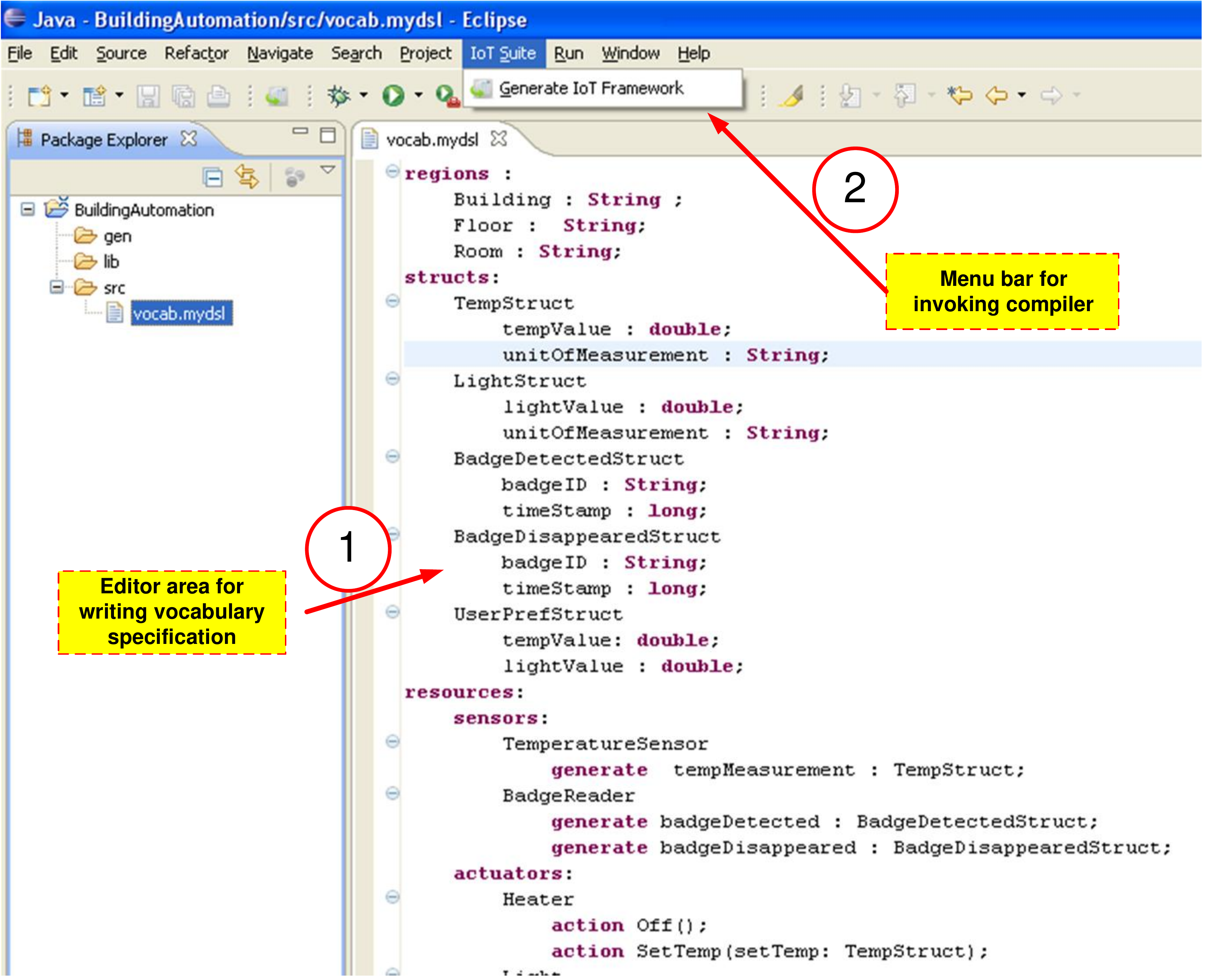} 
	\caption{Editor support for writing vocabulary specification in IoTSuite.} \label{fig:editor}
\end{figure*}

We take an example of building automation domain vocabulary to demonstrate 
an editor support provided by IoTSuite, illustrated in Figure~\ref{fig:editor}. 
The zone \circled{1} shows the editor, where the domain expert writes a domain vocabulary. 
The zone \circled{2}  shows the menu bar, where the domain expert invokes the  
compiler for vocabulary to generate a  framework, a customized 
architecture grammar, and deployment grammar.

\fakeparagraph{\emph{Compiler}} The compiler parses high-level specifications and translates them into code that
can be used by other components in the system. The parser module of compiler is implemented using 
ANTLR\footnote{\url{http://www.antlr.org/}}, 
a well-known parser generator that creates parser files from grammar descriptions. 
The code generator module of compiler manages a repository of plug-ins. 
Each plug-in, defined as template files, is specific to a target implementation language~(e.g., Java, Python). 
The key advantage of it is that it simplifies an implementation of a new code generator for a target implementation. 
The plug-ins are implemented using StringTemplate Engine,\footnote{\url{http://www.stringtemplate.org/}} a Java 
template engine for generating source code or any other formatted text output. We build two compilers 
to aid stakeholders shown in Figure~\ref{fig:devcycle}. (1) compiler for a vocabulary specification, 
and (2) compiler for an architecture specification. The current version of these compilers 
generate frameworks, compatible with Eclipse IDE. 

\begin{figure}[!ht]
	\centering \includegraphics[width=1.0\linewidth]{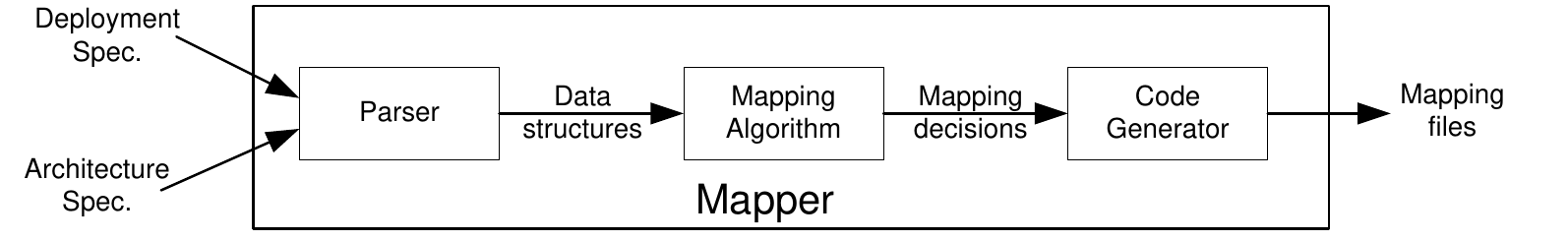} 
	\caption{Architecture of the mapper component in IoTSuite.}\label{fig:mappercomponent}
\end{figure}

\fakeparagraph{\emph{Mapper}}
The mapper produces a mapping from a set of computational services to a set of devices.
Figure~\ref{fig:mappercomponent}  illustrates the architecture of the mapper component. 
The parser converts high-level specifications into appropriate 
data structures that can be used by a mapping algorithm. The mapping algorithm produces mapping decisions 
into appropriate data structures. The code generator consumes the data structures and generates mapping files. 
Our current implementation of the mapper randomly maps computational services to a set of devices. 
However, due to generality of our architecture, more sophisticated mapping algorithm can be plugged into the mapper.

\begin{figure}[!ht]
	\centering \includegraphics[width=1.0\linewidth]{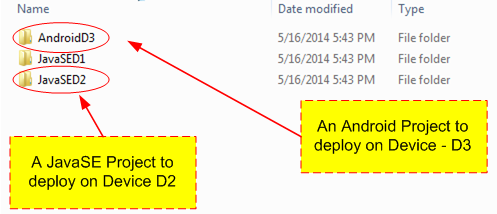} 
	\caption{Packages for target devices compatible with Eclipse IDE} \label{fig:eclipselinker}
\end{figure}

\fakeparagraph{\emph{Linker}}
It combines and packs code generated by various stages of compilation into packages that can be deployed on devices.
The output of the linker is a set of platform-specific projects for devices, specified in the deployment specification. 
These projects are not binaries. They need to be compiled, which can be done by any device-specific 
compiler designed for the target platform. The current version of the linker generates Java source packages 
for Android and JavaSE platform. Figure~\ref{fig:eclipselinker} illustrates packages for 3  target 
devices~(2 packages for JavaSE devices and 1 for Android device), which can be imported into Eclipse IDE. 

\fakeparagraph{\emph{Runtime system}}
The main responsibility of the runtime system is 
a distributed execution of IoT applications.  It is composed of three parts:  
(1) \emph{middleware}: It runs on each individual device and provides 
a support for executing distributed tasks. (2) \emph{wrapper}: It plugs packages, generated 
by the linker module, and middleware. (3) \emph{support library}:  It separates packages,  
produced by the linker component, and underlying middleware by providing interfaces 
that are implemented by each wrapper. The integration of a new middleware into IoTSuite consists 
of an implementation of the following interfaces, specified by the support library, in the wrapper.

\begin{itemize}
	\item \texttt{publish()}. It is an interface for publishing data from a sender.  The definition of this interface 
	    contains: an event name (e.g., temperature), event data (e.g., a temperature value, 
			 Celsius), and publisher's information such as location.
			
   \item \texttt{subscribe()}. It is an interface for receiving event notifications. An interest of events 
	is expressed by sending a subscription request, which contains:
			  a event name~(e.g., temperature), information for filtering events such as regions 
				of interest~(e.g., a RoomAvgTemp component wants to receive events only from 
				a current room), and subscriber's information.
				
    \item \texttt{command()}. It is an interface for triggering an action of an actuator. A command contains: 
		a command name (e.g., switch-on heater), command parameters~(e.g., set temperature of heater to 30$^\circ$C), 
			 and a sender's information.
 
		\item \texttt{request-response()}. It is an interface for requesting data from a requester. In reply, a receiver 
		 sends a response. A request contains a request name (e.g., give profile information), request parameters~(e.g., give 
		  profile of  person with identification 12), and information about the requester.

\end{itemize}

The current implementation of IoTSuite uses the MQTT\footnote{\url{http://mqtt.org/}} middleware,
which enables interactions among Android devices and  JavaSE enabled devices.

%% file: evaluation.tex
\section{Evaluation}\label{chpt:evolution}
The goal of this section is to describe how well the proposed approach addresses our aim in a quantitative manner.  
Unfortunately, the goal is very vague because quality measures are not well-defined and they do not provide a clear 
procedural method to evaluate development approaches in general. We explore development effort, which indicates effort 
required to create an application, that is vital for the productivity of stakeholders~\cite{towardscassou2011}.  

\begin{figure}[!ht]
\centering
\includegraphics[width=0.7\linewidth]{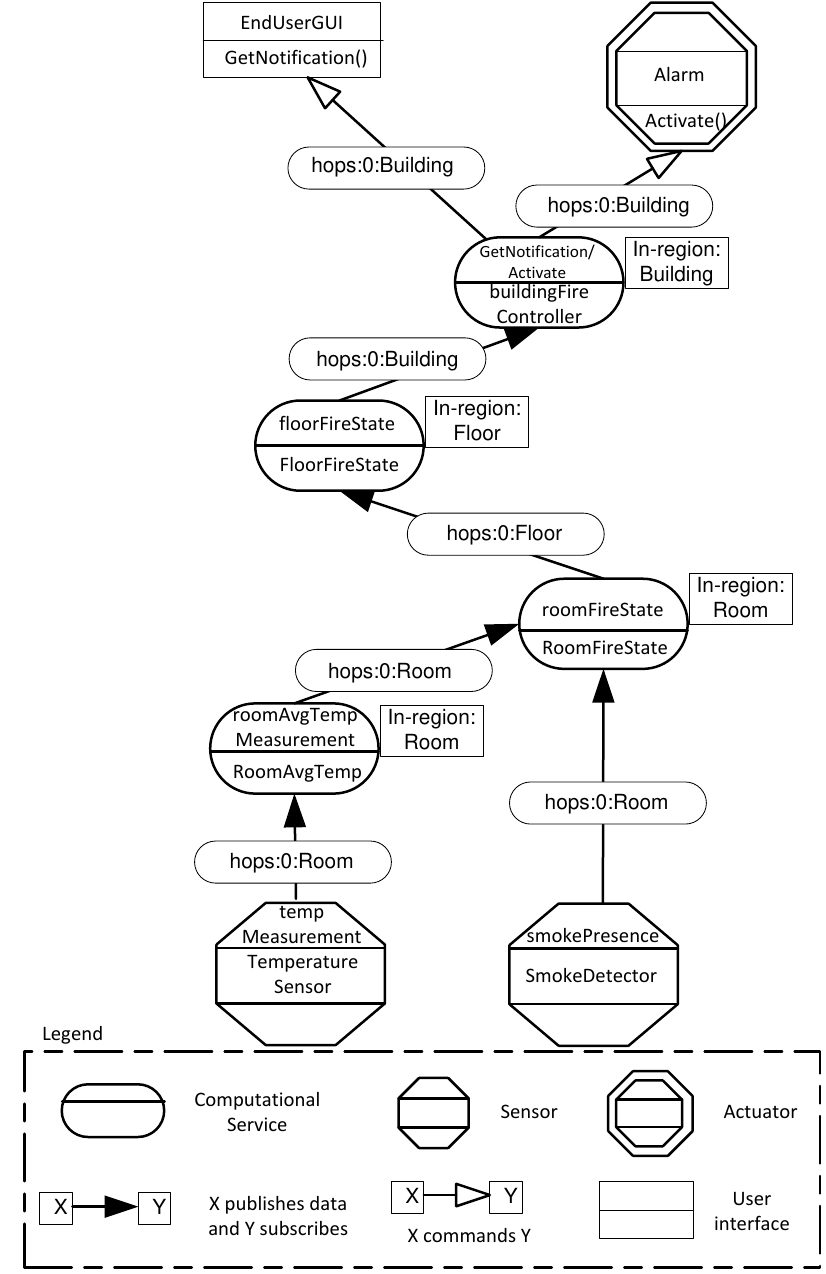}
\caption{Architecture of the fire detection application, similar to work in~\cite{towardscassou2011}.}
\label{fig:firedetect}
\end{figure}

To evaluate our approach we consider two representative IoT applications: (1) the smart building 
application described in Section~\ref{sec:appexample} and (2) a fire detection application, 
which aims to detect fire by analyzing data  from smoke and temperature sensors. When fire occurs, 
residences are notified on their smart phones by an installed application. Additionally, residents 
of the building and neighborhood are informed through a set of alarms. Figure~\ref{fig:firedetect} shows the 
architecture of the fire detection application. A fire state is computed based on a current average temperature value and smoke presence by a local fire state service~(i.e., \texttt{roomFireState}) deployed per room, then a state is sent to a service~(i.e., \texttt{floorFireState}) deployed per floor, and finally a computational service~(i.e., \texttt{buildingFireController}) decides whether alarms should be activated and users should be notified or not.

\subsection{Development effort}\label{sec:developmenteffort}
In order to measure effort to develop an application, we evaluate 
a percentage of a total number of lines of code generated by our approach 
and effort to develop an application involving a large number of devices 
using our approach. we have implemented two IoT applications 
discussed in the previous Section using our approach. These applications are implemented 
independently. We did not reuse specifications and implementations 
of one application in other application. We deployed the two applications on 10 simulated devices 
running on top of a middleware that simulates network on a single PC dedicated to the evaluation.


We measured development effort using Eclipse EclEmma 
2.2.1 plug-in.
This tool counts actual Java statement as lines of 
code and does not consider blank lines or lines 
with comments. Our measurements reveal that more than 82\% of the total 
number of lines of code is generated in  two applications
(see Table~\ref{table:locapp}).

\begin{table*}[!ht]
\centering 
\begin{tabular}{ p{2.0cm} | >{\centering\arraybackslash}p{0.9cm}>{\centering\arraybackslash}p{0.9cm}>{\centering\arraybackslash}p{1.5cm} >{\centering\arraybackslash}p{0.9cm}>{\centering\arraybackslash}p{0.9cm}|>{\centering\arraybackslash}p{1.2cm}>{\centering\arraybackslash}p{1.2cm} >{\centering\arraybackslash}p{1.2cm} | >{\centering\arraybackslash}p{3.2cm} | >{\centering\arraybackslash}p{2.3cm} }
 \toprule
 & \multicolumn{5}{c|}{\scriptsize{\textbf{Handwritten (lines of code)}}} & 
\multicolumn{3}{c|}{\scriptsize{\textbf{Generated (lines of code)}}} & \multicolumn{1}{c|}{\scriptsize{\textbf{\% of Generated code}}} & \\
\scriptsize{Application Name} &	\scriptsize{Vocab Spec.} &	\scriptsize{Arch. Spec.} & \scriptsize{Deploy. Spec.} & \scriptsize{Device driver} & \scriptsize{App. logic} 
& \scriptsize{Mapping code} & \scriptsize{Archi. fram.} & \scriptsize{Vocab. fram.} &  $\frac{generated}{handwritten+ generated}$ & \scriptsize{Code coverage} \\ 
                                   &     &     & \scriptsize{(devices=10)}  &      &   &  &    &  &\\ \midrule
\scriptsize{Smart building}&\scriptsize{41}&\scriptsize{28}&\scriptsize{81}&\scriptsize{98}&\scriptsize{131} & \scriptsize{561}&\scriptsize{408}& \scriptsize{757}&\scriptsize{81.99\%} & \scriptsize{92.22}\\ \midrule
\scriptsize{Fire detection}     & \scriptsize{27} & \scriptsize{21}&\scriptsize{81}&\scriptsize{53}&\scriptsize{72}&\scriptsize{528}&\scriptsize{292}&\scriptsize{476}& \scriptsize{83.61\%} & \scriptsize{90.38}\\ 
\bottomrule
\end{tabular}
\caption{Lines of code in smart building and fire detection applications}  
\label{table:locapp} 
\end{table*}

The measure of lines of code is only useful if the generated code is actually executed. We measured 
code coverage of the generated programming frameworks of two 
applications~(see Table~\ref{table:locapp}) using the
EclEmma\footnote{\url{http://www.eclemma.org/}} Eclipse plug-in. Our
measures show that more than 90\% of generated code is actually
executed, the other portion being error-handling code for errors
that did not happen during the experiment. This high value indicates that most of the execution 
is spent in generated code and that, indeed, our approach reduces development effort by generating
useful code.

The above experiment was conducted for 10 simulated devices. It does not demonstrate  
development effort using our approach for a large number of devices.  Therefore,
the primary aim of this experiment is to evaluate effort to develop an IoT 
application involving a large number of devices. In order to achieve the above aim, 
we have developed the smart building application on a set of simulated device running on top of the middleware 
dedicated to the evaluation. The assessments were conducted over an increasing 
number of devices.  The first development effort assessment was conducted 
on 10 devices instrumented with heterogeneous sensors, actuators, storages, and user 
interfaces. In the next subsequent assessments, 
we kept increasing the number of devices equipped with sensors and actuators.  In each 
assessment, we have measured lines of code to specify vocabulary, architecture, and deployment, 
application logic, and device drivers. Table~\ref{table:locsmartoffice} illustrates 
the assessment results containing a number of devices involved in the experiment and hand-written lines of code 
to develop the smart building application.

\begin{table}[!ht]
\centering 
\begin{tabular}{ >{\centering\arraybackslash}m{2.2cm}  >{\centering\arraybackslash}m{0.9cm}  >{\centering\arraybackslash}m{0.9cm}  >{\centering\arraybackslash}m{1.5cm} >{\centering\arraybackslash}m{0.9cm}  >{\centering\arraybackslash}m{0.9cm} }
 \toprule
 & \multicolumn{5}{c}{\scriptsize{\textbf{Handwritten (lines of code)}}} \\
\scriptsize{Number of  devices} &	\scriptsize{Vocab Spec.} &	\scriptsize{Arch. Spec.} & \scriptsize{Deploy. Spec.} & \scriptsize{Device driver} & \scriptsize{App. logic} \\ \midrule
\scriptsize{10}   &\scriptsize{41}&\scriptsize{28}&\scriptsize{81}&\scriptsize{98}&\scriptsize{131} \\ \midrule
\scriptsize{34}   &\scriptsize{41}&\scriptsize{28}&\scriptsize{273}&\scriptsize{98}&\scriptsize{131}\\ \midrule
\scriptsize{50}   &\scriptsize{41}&\scriptsize{28}&\scriptsize{401}&\scriptsize{98}&\scriptsize{131}\\ \midrule
\scriptsize{62}   &\scriptsize{41}&\scriptsize{28}&\scriptsize{497}&\scriptsize{98}&\scriptsize{131}\\ \midrule
\scriptsize{86}   &\scriptsize{41}&\scriptsize{28}&\scriptsize{689}&\scriptsize{98}&\scriptsize{131}\\ \midrule
\scriptsize{110}  &\scriptsize{41}&\scriptsize{28}&\scriptsize{881}&\scriptsize{98}&\scriptsize{131}\\ \midrule
\scriptsize{200}  &\scriptsize{41}&\scriptsize{28}&\scriptsize{1601}&\scriptsize{98}&\scriptsize{131}\\ \midrule
\scriptsize{300}  &\scriptsize{41}&\scriptsize{28}&\scriptsize{2401}&\scriptsize{98}&\scriptsize{131}\\ \midrule
\scriptsize{350}  &\scriptsize{41}&\scriptsize{28}&\scriptsize{2801}&\scriptsize{98}&\scriptsize{131}\\ \midrule
\scriptsize{500}  &\scriptsize{41}&\scriptsize{28}&\scriptsize{4001}&\scriptsize{98}&\scriptsize{131}\\ \bottomrule
\end{tabular}
\caption{Number of devices involved in the development effort assessment and hand-written lines of 
code to develop the  smart building application.} 
\label{table:locsmartoffice} 
\end{table}
In Table~\ref{table:locsmartoffice}, we have noted the following two observations and their reasons: 
\begin{enumerate}
	\item As the number of devices increases, lines of code for vocabulary and architecture specification, device drivers, 
	and  application logic remain constant for a deployment consisting a large number of devices. 
	The reason is that our approach provides the ability to specify an application at a global level
	rather than individual nodes. 
	\item As the number of devices increases, lines of code for a deployment specification increase. The reason is that the 
network manager specifies each device individually in the deployment specification. This is a limitation of SDL. 
Our future work will be to investigate how a deployment specification  can be expressed in a concise and flexible 
way for a network with a large number of devices. We believe that the use of  regular expressions 
is a possible technique to address this problem.
\end{enumerate}

%% file: relatedwork.tex
\section{Related work}\label{chpt:relatedwork}

This Section focuses on existing works in literature that would address the research 
challenges discussed in Section~\ref{sec:challenges}. As stated earlier, while the application development 
life-cycle has been discussed in general in the software engineering domain, a similar structured 
approach is largely lacking in the IoT for the development of Sense-Computer-Control~(SCC)~\cite[p.~97]{softwareArchtaylor2010}  
applications. Consequently, in this Section we present existing approaches geared towards the IoT, 
but also its precursor fields of Pervasive Computing and Wireless Sensor Networking. 
These are mature fields, with several excellent surveys available on 
programming models~\cite{programmingSurveysugihara2008, programmingWSNmottola2011} 
and middleware~\cite{henricksen2006survey}. 

We organize this Section based on the perspective of the system provided to the stakeholders by the various approaches. 
Section~\ref{sec:nodelevelapproach} presents the node-level programming approaches, where the developer 
has significant control over the actions of each device in the system, which comes at the cost of complexity. 
Section~\ref{sec:db} summarizes approaches that aim to abstract the entire (sensing) system as a database 
on which one can run queries. Section~\ref{sec:macro} presents the evolution of these approaches 
to macroprogramming inspired by general-purpose programming languages, where abstractions 
are provided to specify high-level collaborative behaviors at the system-level while hiding 
low-level details from stakeholders. Section~\ref{sec:mde} then describes the macroprogramming 
approaches more grounded in model-driven development techniques, which aim to provide a cleaner 
separation of concerns during the application development process. We summarize all these approaches 
in Table~\ref{summarytable1}.

\subsection{Node-centric programming}\label{sec:nodelevelapproach}
In the following, we present systems that adopt the node-centric approach.

In the pervasive computing domain, Olympus~\cite{olympusranganathan2005} is a programming model on top of 
Gaia~\cite{roman2002gaia} -- a distributed middleware infrastructure for 
pervasive environments. Stakeholders write a C++ program that consists of a high-level description 
about active space entities (including service, applications, devices, physical 
objects, and locations) and common active operations~(e.g., switching devices on/, 
starting/stopping applications).  The Olympus framework takes 
care of resolving  high-level description  based on properties specified by stakeholders. 
While this approach certainly simplifies the SCC application development involving heterogeneous devices, 
stakeholders have to write a lot of code to interface hardware and software components, as well 
as to interface software components and its interactions with a distributed system. 
This makes it tedious to develop applications involving a large number of devices. 

The Context toolkit~\cite{contexttoolkitdey2001,salber1999context} simplifies the context-aware 
application development on top of heterogeneous data sources by providing three architectural 
components, namely, widgets, interpreters, and aggregators. These components separate application 
semantics from platform-specific code. For example, an application does not have to be modified 
if an Android-specific sensor is used rather than a Sun SPOT sensor. It  means stakeholders can 
treat a widget in a similar fashion and do not have to deal with differences among platform-specific 
code. Although context toolkit provides support for acquiring the context data from the 
heterogeneous sensors, it does not support actuation that is an essential part 
of IoT applications.

Henricksen et al.~\cite{MDDhenricksen2006, bettini2010survey} propose  a middleware  
and a programming framework to gather, manage, and disseminate context to  applications.  
This work introduces context modeling concepts, namely, context modeling languages, 
 situation abstraction; and preference and branching models. This work presents 
a software engineering process that can be used in conjunction with the specified concepts. 
However,  the clear  separation of  roles among the various stakeholders is missing. Moreover, 
this framework limits itself to context gathering applications, thus not  
providing the actuation support that is important for IoT application development.

\fakeparagraph{Physical-Virtual mashup} As indicated by its name, it connects web services from both the 
physical and virtual world through visual constructs directly from web browsers.  The embedded device runs 
tiny web servers~\cite{duquennoy2009web} to answer HTTP queries from users for checking or changing the 
state of  a device. For instance, users may want to see temperature of different places on map. Under 
such requirements, stakeholders can use the mashup to connect physical services such as temperature sensors 
and virtual services such as Google map.  Many mashup prototypes have been developed that include both the 
physical and virtual services~\cite{blackstock2012wotkit, guinard2010resource, castellani2012webiot, ghidini2012fuseviz, 
gupta2010early}. The mashup editor usually  provides visual components representing web service and operations (such as add, filter) 
that stakeholders need to connect together to program an application. The framework takes care of resolving these visual 
components based on properties specified by stakeholders and produces code to interface software components and distributed system.
The main advantage of  this mashup approach is that any service, either physical or virtual, can be mashed-up if they follow the standards~(e.g., REST). 
The Physical-Virtual mashup significantly lowers the barrier of the application development. 
However, stakeholders have to manage a potentially large graph 
for an application involving a large number of entities. This makes it difficult 
to develop applications containing a large number of entities.

\subsection{Database approach}\label{sec:db}
In TinyDB~\cite{madden2005tinydb} and  Cougar~\cite{yao2002cougar} systems, an SQL-like query  
is submitted to a WSN. On receiving  a query, the system collects data from the individual 
device, filters it, and sends it to the base station. They provide a suitable interface 
for data collection in a network with a large number of devices. However,  
they do not offer much flexibility for introducing the application logic. For example, stakeholders
require extensive modifications in the TinyDB parser and query engine to implement new query operators. 

The work on SINA~(Sensor Information Networking Architecture)~\cite{SINAmiddlewareshen2001} overcomes this limitation on specification of custom operators
by introducing an imperative language with an SQL query. In SINA, stakeholders can embed a script 
written in Sensor Querying and Tasking Language~(SQTL)~\cite{jaikaeo2000querying} in the SQL query. By this hybrid 
approach, stakeholders can perform more collaborative tasks than what SQL in TinyDB and Cougar can describe. 

The TinyDB, Cougar, and SINA systems are largely limited to homogeneous devices. 
The IrisNet~(Internet-Scale Resource-Intensive Sensor Network)~\cite{gibbons2003irisnet} allows stakeholders to query a large 
number of distributed heterogeneous devices. For example, Internet-connected PCs source sensor feeds and cooperate to answer queries. Similar to the other database 
approaches, stakeholders view the sensing network as a single unit that supports a high-level query 
in XML. This system provides a suitable interface for data collection from a large number 
of different types of devices. However, it does not offer flexibility for introducing 
the application logic, similar to TinyDB and Cougar.  

Semantic Streams~\cite{whitehouse2006semantic} 
allows stakeholders to pose a declarative query over semantic interpretations of sensor data. For example, 
instead of querying raw magnetometer data, stakeholders query whether a vehicle is a car or truck. The system 
infers this query and decides sensor data to use to infer the type of vehicle. The main benefit of using 
this system is that it allows people, with less technical background to query the network with heterogeneous devices. 
However, it presents a centralized approach for sensor data collection that limits its applicability 
for handling a network with a large number of devices.

\fakeparagraph{Standardized protocols-based systems}
A number of systems have been proposed to expose functionality of devices accessible 
through standardized protocols without having worry about the heterogeneity of underlying 
infrastructure~\cite{mohamed2011survey}. They logically view sensing devices~(e.g., motion sensor, 
temperature sensor, door  and window sensor) as service providers for applications and provide 
abstractions  usually through  a set of services. We discuss these examples below.

Priyantha et al.~\cite{priyantha2008tiny} present an approach based on SOAP~\cite{box2000simple} to enable an 
evolutionary WSN where additional devices may be added after the initial deployment. To support 
such a system, this approach has adopted two features. (1) structured data: the data generated 
by sensing devices are represented in a XML format  for that may be understood by any application. 
(2) structured functionality: the functionality of 	a sensing device is exposed by Web Service 
Description Language (WSDL)~\cite{chinnici2007web}. While this system addresses the evolution issue 
in a target deployment, the authors do not demonstrate  the evolution scenarios such as a change in 
functionality of an application, technological advances in deployment devices.

A number of approaches based on REST~\cite{fielding2000architectural} have been proposed to overcome the resource needs and 
complexity of SOAP-based web services for sensing and actuating devices. TinyREST~\cite{luckenbach2005tinyrest} is one 
of first attempts to overcome these limitations. It uses the HTTP-based REST architecture to 
access a state of sensing and actuating devices. The TinyREST gateway maps the HTTP request 
to TinyOS messages and allows stakeholders to access sensing and actuating devices 
from their applications. The aim of this system is to make services available through standardized 
REST without having to worry about the heterogeneity of the underlying infrastructure; that said, it 
suffers from a centralized structure similar to TinySOA.

\subsection{Macroprogramming languages}\label{sec:macro} 
In the following, we present macroprogramming languages 
for IoT application development, which are grounded in traditional general 
purpose programming languages (whether imperative or functional) in order 
to provide developers with familiar abstractions.

Kairos~\cite{macrogummadi2005} allows stakeholders to program an application 
in a Python-based language. The Kairos developers write a 
centralized program of a whole application.  Then, the pre-processor divides the program into subprograms, 
and later its compiler compiles it into binary code containing code for accessing local and remote variables.  
Thus, this binary code allows stakeholders to program distributed sensor network applications.  Although  
Kairos makes the development task easier for  stakeholders, it targets homogeneous network where each device 
executes the same application.

Regiment~\cite{regimentnewton2007} provides a high-level  programming language based on Haskell to describe an application as a set of spatially distributed data streams. 
This system provides primitives that facilitate processing data, manipulating regions, and 
aggregating data across regions. The written program is compiled down to an 	intermediate token 
machine language that passes information over a spanning tree constructed across 	the WSN. 
In contrast to the database approaches, this approach provides greater flexibility to stakeholders when it 
comes to the application logic. However, the regiment program collects 
data to a single base station. It means that the flexibility for  any-to-any device collaboration for 
reducing scale is difficult. 

MacroLab~\cite{hnat2008macrolab} offers a vector programming abstraction similar to Matlab for applications 
involving both sensing and actuation. Stakeholders write a single program for an entire application using Matlab 
like operations such as \texttt{addition}, \texttt{find}, and \texttt{max}. The written macroprogram is 
passed to the MacroLab decomposer that generates multiple decompositions of the program. Each decomposition 
is analyzed by the cost analyzer that calculates the cost of each decomposition with respect to a cost profile (provided by 
stakeholders) of a target deployment. After choosing a best decomposition by the cost analyzer, it is passed 
to the compiler that converts the decomposition into a binary executable. The main benefit is that 
it offers flexibility of decomposing code according to cost profiles of the target platform. 
While this system certainly separates the deployment aspect and functionality of an application, 
this approach  remains  general purpose and provides little guidance to stakeholders about 
the application domain.

\subsection{MDD approach}\label{sec:mde} 
A number of model-driven approaches have been proposed to make IoT application development easy, described below. 

PervML~\cite{pervMLserral2010} allows stakeholders to specify pervasive  applications 
 at a high-level of abstraction using a set of models.  This system raises the level 
of abstraction in program specification, and code generators produce code from these 
specifications.  Nevertheless, it adopts generic UML notations to describe them, thus 
provides little guidance to stakeholders about the specific application domain.  In addition 
to this, the main focus of this work is to address the heterogeneity associated with 
pervasive computing applications, and the consideration of a large number of devices 
in an application is missing.  PervML integrates the mapping process at the deployment phase.  
However, stakeholders have to link the application code and configure device drivers manually.  
This manual work in the deployment phase is  not suitable for IoT applications involving 
a large number of devices.  Moreover, the separation between deployment and domain-specific 
features are missing.  These limitations would restrict PervML to a certain level.


DiaSuite~\cite{towardscassou2011} is a suite of tools to develop pervasive computing applications. 
It combines design languages and covers application development life-cycle. 
The design language defines both a taxonomy of an application domain and an application architecture. 
Stakeholders define entities in a high-level manner to 
abstract heterogeneity. However, the consideration of a large number of devices in an application 
is largely missing. Moreover, the application deployment for a large number of heterogeneous devices using this approach 
is difficult because stakeholders require manual effort (e.g., mapping of computational services 
to devices). 

ATaG~\cite{pathakhal00723799}, which is a WSN is a macroprogramming framework to develop 
SCC applications. ATaG presents a compilation framework that 
translates a program, containing abstract notations, into executable node-level programs. Moreover, it tackles 
the issue of scale reasonably well. The ATaG linker and mapper modules support 
the application deployment phase by producing device-specific code to result 
in a distributed software system collaboratively hosted by individual devices, 
thus providing automation at deployment phase. Nevertheless, the clear 
separation of  roles among the various stakeholders in the application 
development, as well as the focus on heterogeneity among the 
constituent devices are largely missing. Moreover, the ATaG program notations 
remains general purpose and provides little guidance to stakeholders about 
the application domain. 

RuleCaster~\cite{bischoff2006rulecaster,bischoff2007life} introduces an engineering 
method to provide  support for SCC applications,  as well as 
evolutionary changes in  the application development. The RuleCaster programming model 
is based on a logical partitioning of the network into spatial regions. The RuleCaster compiler 
takes as input the application program containing rules 
and a network model that describes device locations and its capabilities.  Then, it 
maps processing tasks to devices.  Similar to ATaG, this system handles the scale 
issue reasonably well by partitioning the network into several spatial regions.  
Moreover, it supports automation at the deployment phase by mapping computational
components to devices.  However, the clear separation of roles among the various 
stakeholders, support for application domain, as well as the focus on 
heterogeneity among the constituent devices are missing. 

Pantagruel~\cite{drey2009taxonomy} is a visual approach dedicated to the development 
of home automation applications.  The Pantagruel application development consists of three steps: 
(1) specification of taxonomy to define entities of the home automation domain~(e.g., temperature sensor, alarm, 
door, smoke detector, etc.),  (2) specification of rules 
to orchestrate these entities using the Pantagruel visual language, and (3) compilation of 
the taxonomy and orchestration rules to generate a programming framework. The novelty of this 
approach is that the orchestration rules are customized with respect to entities defined in the taxonomy. 
While this system reduces the requirement of having domain-specific knowledge for  other stakeholders, 
the clear separation of different development concerns, support for large scale, automation both at the development 
and deployment phase are largely missing. These limitations make it
difficult to use for IoT application development.

\newcolumntype{A}{>{\columncolor[gray]{0.8}[\tabcolsep][\tabcolsep]}c}
\newcolumntype{B}{>{\columncolor[gray]{0.95}} c|}
\begin{table*}[!ht]
\centering
\begin{tiny}
\renewcommand{\arraystretch}{1.5} 
\onecolumn
\begin{longtable}{c l p{3.5cm} >{\centering\arraybackslash}m{1.8cm}  >{\centering\arraybackslash}m{1cm} >{\centering\arraybackslash}m{1cm} >{\centering\arraybackslash}m{2cm} >{\centering\arraybackslash}m{1.2cm} >{\centering\arraybackslash}m{1.2cm} >{\centering\arraybackslash}m{1.5cm} >{\centering\arraybackslash}m{1.5cm} >{\centering\arraybackslash}m{3cm}} \toprule

\multicolumn{2}{c}{Existing Approaches} &  &  Division of roles&Hetro.&Scale&\multicolumn{3}{c}{Life-cycle phases}\\\cmidrule{7-9}

\multicolumn{2}{c}{}                    &  &                   &      &     & Development phase  &  Deployment phase& Maintenance phase
\\\toprule
\endfirsthead
						
							 &                &\parbox{3cm}{ContextToolkit~\cite{contexttoolkitdey2001}~(2001)} &$\times$ &$\sim$   &$\times$ &$\times$ &$\times$ &$\sim$ \\\cmidrule{3-12}
	             &  Node-centric  &Olympus~\cite{olympusranganathan2005}~(2005)        &$\times$ &$\surd$  &$\times$ &$\times$ &$\times$ &$\times$\\\cmidrule{3-12}						
							 &  Programming   &\parbox{3cm}{Henricksen et al.~\cite{bettini2010survey}~(2010)}&$\times$&$\sim$&$\times$&$\sim$&$\times$&$\times$\\\cmidrule{3-12}   
							 &                &\parbox{3cm}{Dominique et al.~\cite{guinard2010resource}~(2010)}&$\times$&$\surd$&$\times$&$\sim$&$\times$&$\sim$ \\\cmidrule{1-12}		
							 &      				  &TinyDB~\cite{madden2005tinydb}~(2000)& $\times$ &$\times$&$\sim$&Not clear & $\times$&$\times$ \\\cmidrule{3-12}  
               &      				  &IrisNet~\cite{gibbons2003irisnet}~(2000)& $\times$&$\sim$&$\sim$&Not clear&$\times$&$\times$ \\\cmidrule{3-12} 
               &  Database      &SINA~\cite{SINAmiddlewareshen2001}~(2000)&$\times$&$\times$&$\surd$&Not clear&$\times$&$\times$\\\cmidrule{3-12}  							
							 &  Approach              &TinyREST~\cite{luckenbach2005tinyrest}~(2005)&$\times$&$\surd$&$\times$&$\times$&$\times$&$\sim$ \\\cmidrule{3-12}
							 &                &\parbox{3cm}{Semantic Streams~\cite{whitehouse2006semantic}~(2006)}&$\times$&$\sim$&$\times$&$\sim$&$\times$&$\times$\\\cmidrule{3-12}			
							 &              &\parbox{3cm}{Priyantha et al.~\cite{priyantha2008tiny}~(2008)}&$\times$&$\surd$&$\times$&$\times$&$\times$&$\sim$\\ \cmidrule{1-12}		
               & Macroprogramming   &Kairos~\cite{macrogummadi2005}~(2005)      &$\times$ &$\times$ &$\surd$ &$\sim$ &$\sim$    &$\times$ \\\cmidrule{3-12}
               &  Languages         &Regiment~\cite{regimentnewton2007}~(2007)    &$\times$ &$\times$ &$\sim$  &$\sim$ &$\sim$    &$\times$\\\cmidrule{3-12}					  
               &                    &MacroLab~\cite{hnat2008macrolab}~(2008)    &$\times$ &$\times$ &$\surd$ &$\sim$ &$\surd$   &$\sim$ \\\cmidrule{1-12}

							 &                    &RuleCaster~\cite{bischoff2007life}~(2007)&$\times$  &$\times$ &$\surd$  &$\sim$  &$\sim$    &$\sim$\\ \cmidrule{3-12}							
							 &  Model-driven      &Pantagruel~\cite{drey2009taxonomy}~(2009)&$\sim$    &$\sim$   &$\times$ &$\sim$  &$\times$  &$\sim$\\ \cmidrule{3-12}							
							 &  Development       &PervML~\cite{pervMLserral2010}~(2010)    &$\sim$    &$\surd$  &$\times$ &$\sim$  &$\sim$    &$\sim$\\ \cmidrule{3-12}
               &                    &DiaSuite~\cite{towardscassou2011}~(2011)  & $\sim$   &$\surd$  &$\times$ &$\surd$ &$\times$  &$\sim$\\ \cmidrule{3-12}
	             &                    &ATaG~\cite{pathakhal00723799}~(2011)      & $\times$ &$\sim$   &$\surd$  &$\sim$  &$\surd$   &$\sim$\\ \cmidrule{1-12} \\
\caption{Comparison of existing approaches. $\surd$ -- Supported, $\times$ -- No supported, $\sim$ -- No adequately supported.}
\label{summarytable1}
\end{longtable}
\twocolumn
\end{tiny}
\end{table*}

%% file: conclusion.tex
\section{Conclusion}\label{chapt:conclusionandfuturework}

This paper presents a development methodology for IoT application development, based on techniques 
presented in the domains of sensor network macroprogramming and model-driven development.  
It separates IoT application development into different concerns and integrates a set 
of high-level languages to specify them. This approach is supported by automation techniques at different 
phases of IoT application development and allows an iterative development to handle 
evolutions in different concerns. Our evaluation based on two realistic 
IoT applications shows that our approach generates a significant percentage of the total application code, 
drastically reduces development effort for IoT applications involving a large number of devices.
Our approach addresses the challenges discussed in Section~\ref{sec:challenges} in the following manner:

\fakeparagraph{\emph{Lack of division of roles}} 
Our approach identifies roles of each stakeholder and separates them according 
to their skills. The clear identification of expectations and specialized skills of each 
stakeholder helps them to play their part effectively, thus promoting a suitable 
division of work among stakeholders involved in IoT application development. 

\fakeparagraph{\emph{Heterogeneity}} 
SAL and SVL provide abstractions to specify different types of devices, 
as well as heterogeneous interaction modes in a high-level manner. Further, 
high-level specifications written using SAL and SVL are compiled to a programming 
framework that (1) abstracts heterogeneous interactions among software components 
and (2) aids the device developers to write code for different 
platform-specific implementations. 

\fakeparagraph{\emph{Scale}}
SAL allows the software designer to express his requirements in a compact manner 
regardless of the scale of a system.  Moreover, it offers scope constructs to 
facilitate scalable operations within an application.  They reduce scale by 
enabling hierarchical clustering in an application. To do so, these constructs 
group devices to form a cluster based on their spatial relationship (e.g., ``devices are in room\#1'').  
Within a cluster, a cluster head is placed to receive and process data from 
its cluster of interest. The grouping could be recursively applied 
to form a hierarchy of clusters. The scale issue is thus handled, thanks to the use of a middleware that supports logical scopes and regions.

\fakeparagraph{\emph{Different life cycle phases}}
Our approach is supported by code generation, task-mapping, and linking techniques. 
These techniques together provide automation at different life cycle phases. At the 
development phase, the code generator produces (1) an architecture framework that 
allows the application developer to focus on the application logic 
by producing code that hide low-level interaction details and 
(2) a vocabulary framework to aid the device developer to implement platform-specific device drivers.  
At the deployment phase, the mapping and linking together produce device-specific code 
to result in a distributed software system collaboratively hosted by individual devices. 
To support maintenance phase, our approach separates IoT application development into different 
concerns and allows an iterative development, supported by the automation techniques.

\section{Future work}\label{sec:futurework}

This paper addresses the challenges, presented by the steps involved in IoT application development, 
and prepares a  foundation  for  our future research work. Our future work will proceed in the 
following complementary directions, discussed below. 

\fakeparagraph{\emph{Mapping algorithms cognizant of heterogeneity}}
While the notion of region labels is able to
        reasonably tackle the issue of scale at an abstraction level,
        the problem of heterogeneity among the devices still
        remains. We will provide rich abstractions to express both the
        properties of the devices (e.g., processing and storage
        capacity, networks it is attached to, as well as monetary cost
        of hosting a computational service), as well as the requirements
        from stakeholders regarding the preferred placement of the
        computational services of the applications. These will then be
        used to guide the design of algorithms for efficient mapping
        (and possibly migration) of computational services on devices.

\fakeparagraph{\emph{Developing concise notion for SDL}}
In the current version of SDL, the network manager is forced to specify the detail 
of each device individually. This approach works reasonably  well  in a target deployment
 with a small number of devices. However, it may be time-consuming and error-prone  
for a target deployment consisting of  hundreds to thousands of devices. Our future  work will be 
to investigate how the deployment specification  can be expressed in a concise and flexible 
way for a network with a large number of device. We believe that the use of regular 
expressions is a possible technique to address this problem.

\fakeparagraph{\emph{Testing support for IoT application development}}
Our near term future work will be  to provide support for the testing phase. A key advantage of 
testing is that it emulates the execution of an application before deployment so as to identify 
possible conflicts, thus reducing application debugging effort. The support will be provided 
by integrating an open source simulator. This simulator will enable transparent 
testing of IoT applications in a simulated physical environment. Moreover, we expect to enable the simulation of 
a hybrid environment, combining both real and physical entities. Currently, we are investigating 
open source simulators for IoT applications. We see Siafu\footnote{\url{http://siafusimulator.org/}} 
as a possible candidate due to its open source and thorough documentation.

\fakeparagraph{\emph{Run-time adaptation in IoT applications}}
 Even though our approach addresses the challenges posed by evolutionary 
			 changes in target deployments and application requirements, stakeholders 
			 have to still recompile the updated code.  This is common practice in a 
			 single PC-based development environment, where recompilation is generally 
			 necessary to integrate changes.   However, it would be very interesting to 
			 investigate how changes can be injected into the running application that 
			 would adapt itself accordingly. For instance, when a new device is added into 
			 the target deployment, an IoT application can autonomously include a new device and 
			 assign a task that contributes to the execution of the currently running application.